\documentclass[11pt]{article}

\usepackage{algorithmic} 
\usepackage{algorithm} 
\usepackage{setspace} 
\usepackage{rotating} 
\usepackage[margin=1in]{geometry} 
\usepackage{amsmath} 
\usepackage{amsthm} 
\usepackage{amsfonts, amssymb} 
\usepackage{graphicx} 
\usepackage{epstopdf} 
\usepackage{hyperref} 
\usepackage{bookmark}
\usepackage{enumerate} 
\usepackage{bm} 
\usepackage{authblk} 
\usepackage{multirow} 
\usepackage{latexsym} 
\usepackage{natbib}
\usepackage{rotating} 
\usepackage{titlesec} 
\usepackage{caption} 
\usepackage{mathtools} 
\usepackage{color} 
\usepackage{cprotect}

\hypersetup{pdfauthor={Name}}

\makeatletter 
\renewcommand\@biblabel[1]{#1} 
\makeatother

\newtheorem{theorem}{Theorem}

\theoremstyle{definition}						
\newtheorem{lem}{Lemma}				

\theoremstyle{definition}

\theoremstyle{definition}
\newtheorem{assumption}{Assumption}

\theoremstyle{definition}

\theoremstyle{definition}

\def\T{{ \mathrm{\scriptscriptstyle T} }}

\newcommand{\hpi}{\widehat{\gamma}}
\newcommand{\Sphi}{\widehat{\Sigma}_{\Gamma}}
\newcommand{\Spi}{\widehat{\Sigma}_{\gamma}}

\newcommand{\inP}{\stackrel{p}{\to}}
\newcommand{\reals}{\mathbb{R}}

\begin{document}
\setstretch{1.54}

\title{Weak-Instrument Robust Tests in Two-Sample Summary-Data Mendelian Randomization}
\author{Sheng Wang\thanks{shengw@stat.wisc.edu} \ and Hyunseung Kang\thanks{hyunseung@stat.wisc.edu}\\ \small{\textit{Department of Statistics, University of Wisconsin-Madison}}}
\date{}
\maketitle
\begin{abstract}
Mendelian randomization (MR) \textcolor{black}{has been} a popular method in genetic epidemiology to estimate the effect of an exposure on an outcome using genetic variants as instrumental variables (IV), with two-sample summary-data MR being the most popular. Unfortunately, instruments in MR studies are often weakly associated with the exposure, which can bias effect estimates and inflate Type I errors. In this work, we propose test statistics that are robust under weak instrument asymptotics by extending the Anderson-Rubin, Kleibergen, and the conditional likelihood ratio test in econometrics to two-sample summary-data MR. We also use the proposed Anderson-Rubin test to develop a point estimator and to detect invalid instruments. We conclude with a simulation and an empirical study and show that the proposed tests control size and have better power than existing methods with weak instruments.


\end{abstract}

\setstretch{1.34}

\section{Introduction}
Mendelian randomization (MR) \textcolor{black}{has been} a popular method in genetic epidemiology to study the effect of modifiable exposures on health outcomes using genetic variants as instrumental variables (IV). In a nutshell, MR uses instruments, typically single nucleotide polymorphisms (SNPs), from publicly available genome-wide association studies (GWAS). These instruments must be (A1) associated with the exposure, (A2) independent of unmeasured confounders, and (A3) independent of the outcome variable after conditioning on the exposure \citep{davey2003mendelian,lawlor2008mendelian}. Typically, two non-overlapping GWAS are used to find instruments, one GWAS studying the exposure and another GWAS studying the outcome. Also, due to privacy, when estimating the exposure effect, only summary statistics instead of individual-level data are used. This setup is commonly known as two-sample summary-data MR \citep{pierce2013efficient, burgess2013mendelian, burgess2015using}.

The focus of this paper is on robust inference of the exposure effect in two-sample summary-data MR when (A1) is violated, or when the instruments are weakly associated with the exposure in the form of weak IV asymptotics \citep{staiger65stock}; see Section \ref{review} for details. Many genetic instruments in MR studies only explain a fraction of the variation in the exposure and using these weak instruments can introduce bias and inflate Type I errors in MR studies
\citep{burgess2011bias}. Weak IVs can also amplify bias from minor violations of (A2) and (A3) \citep{small2008war} and as such, using weak-IV robust methods may not only reduce bias from weak instruments, but also attenuate biases from invalid instruments. {\color{black} Historically, MR methods assumed instruments are strongly associated with the exposure. 
For example, methods such as the inverse-variance weighted estimator (IVW) \citep{burgess2013mendelian}, MR-Egger regression (MR-Egger) \citep{bowden2015mendelian}, the weighted median estimator \citep{bowden2016consistent} and the modal estimator \citep{hartwig2017robust} often assumed that each instrument's correlation to the exposure is measured without error. Recently, \citet{ye2019debiased} showed that confidence intervals from the IVW estimator will have lower-than-expected coverage under weak instruments. Also, \citet{bowden2016assessing}, \citet{bowden2017framework}, \citet{bowden2018improving}, and \citet{bowden2019improving} proposed more robust methods, notably the exact IVW estimator in Box 2 of \citet{bowden2019improving} and the robust MR-Egger estimator via radial regression in \citet{bowden2018improving}. But, these robust methods have not been shown to satisfy a key necessary requirement for a valid confidence interval, where the confidence interval becomes infinite  with at least probability $1-\alpha$ when faced with weak IVs in order to maintain $1-\alpha$ coverage  \citep{dufour1997some}.}

Many works in econometrics have dealt with weak instruments; see \citet{stock2002survey} for an overview. In particular, the Anderson-Rubin (AR) test \citep{anderson1949estimation}, the Kleibergen (K) test \citep{kleibergen2002pivotal}, and the conditional likelihood ratio (CLR) test \citep{moreira2003conditional} provide Type I error control regardless of instrument strength and they satisfy the aforementioned necessary requirement for valid confidence intervals. However, all these methods assume that individual-level data is available and the data comes from the same sample. As mentioned earlier, in MR, one rarely has access to individual-level data due to privacy and is forced to work with anonymized summary statistics from multiple GWAS.


Our contribution is to extend the three aforementioned weak-instrument robust tests in econometrics, the AR test, the K test , and the CLR test, to work with two-sample summary data. In particular, we leverage \textcolor{black}{recent work} by \citet{choi2018weak} who worked with two-sample, but individual data, and extend it to summary-data settings. We show that  these modified tests, which we call mrAR, mrK, and mrCLR, have asymptotic size control. {\color{black} We also show that mrAR can be used to recover estimators proposed by \citet{bowden2019improving} and \citet{zhao2020statistical}}
and to detect presence of invalid instruments, \textcolor{black}{similar to the test for pleiotropy using the Q statistic in \citet{bowden2019improving}}. We conclude with empirical studies and  \textcolor{black}{provide recommendations on how to use our methods in practice.}

\section{Setup and Method}
\subsection{Review: Two-Sample Summary Data in MR}\label{review}
We review the data generating model underlying MR. Suppose we have two independent groups of people, with $n_1$ and $n_2$ participants \textcolor{black}{in} each of the two groups. For each individual $i$ in sample $l = 1,2$, let $Y_{li} \in \reals$ denote his/her outcome, $D_{li} \in \reals$ denote his/her exposure, and $Z_{li} \in \reals^{L}$ denote his/her $L$ instruments. Single-sample individual-data MR assumes that for one sample $l \in \{1,2\}$, $Y_{li}, D_{li}, Z_{li}$ \textcolor{black}{follow} a linear structural model. 
\begin{align} \label{eq:model_iv}
\begin{split}
&\ Y_{li} = \beta_{\rm int} + D_{li} \beta + \epsilon_{li}, \quad{} D_{li} = \gamma_{\rm int} + Z_{li}^\intercal \gamma+ \delta_{li}, \quad{} E[\epsilon_{li}, \delta_{li} \mid Z_{li}] = 0
\end{split}
\end{align}
The parameter of interest is $\beta$ and has a causal interpretation under some assumptions \citep{holland1988causal, kang2016instrumental, zhao2017two}. Two-sample individual-data MR assumes the same underlying structural model \eqref{eq:model_iv} for both samples. But, for sample $l=1$, the investigator only sees $(Y_{1i}, Z_{1i})$ and for sample $l=2$, the investigator only sees  $(D_{2i}, Z_{2i})$ \citep{pierce2013efficient, burgess2013mendelian}. 
Finally, in two-sample summary-data MR, only summarized statistics of $(Y_{1i}, Z_{1i})$ and $(D_{2i}, Z_{2i})$ are available. Specifically, from $n_1$ samples of $(Y_{1i}, Z_{1i})$, we obtain (i) $\widehat{\Gamma} \in \reals^L$ where $\widehat{\Gamma}_{j}$ is the estimated association between IV $Z_{1ij}$ and $Y_{1i}$ and (ii) $\Sphi \in \reals^{L \times L}$, the estimated covariance of $\widehat{\Gamma}$. Similarly, from $n_2$ samples of $(D_{2i}, Z_{2i})$, we obtain (i) $\widehat{\gamma} \in \reals^L$ where $\widehat{\gamma}_j$ is the estimated association between IV $Z_{2ij}$ and $D_{2i}$ and (ii) $\Spi \in \reals^{L \times L}$, the estimated covariance of $\widehat{\gamma}$. We assume that the summary statistics $(\widehat{\Gamma}, \widehat{\Sigma}_{\Gamma}, \widehat{\gamma}, \widehat{\Sigma}_{\gamma})$ used in the analysis satisfy the following assumptions:
\begin{assumption} \label{as:1} The IV-exposure and the IV-outcome effects are independent, $\widehat{\gamma} \perp \widehat{\Gamma}$.
\end{assumption}
\begin{assumption} \label{as:2} The two effect estimates follow $\widehat{\gamma} \sim N(\gamma, \Sigma_{\gamma})$ and $\widehat{\Gamma} \sim N(\gamma \beta, \Sigma_{\Gamma})$.
\end{assumption}
\begin{assumption} \label{as:3} We have  $n_1 (\Sphi - \Sigma_{\Gamma}) \inP 0$, $n_2 (\Spi - \Sigma_{\gamma}) \inP 0$ and $n_1 \Sigma_{\Gamma} \inP \Sigma_1, n_2 \Sigma_{\gamma} \inP \Sigma_2,$ where $\Sigma_1, \Sigma_2$ are deterministic, positive-definite matrices. 
\end{assumption}
\begin{assumption} \label{as:4} For some constant $C \in \reals^L$, we have $\gamma = C/\sqrt{n_2}$.
\end{assumption}
Assumption \ref{as:1} is typically satisfied in MR by having two GWAS that independently measure SNPs' associations to the exposure and the outcome \citep{pierce2013efficient}. Assumption \ref{as:2} is reasonable in publicly-available GWAS where the effect estimates are based on running ordinary least squares (OLS) 
and the sample size is typically on the order of thousands for the central limit theorem to hold. \textcolor{black}{Assumption \ref{as:3} states that the estimated covariance matrices converge to the true covariance matrices and the true covariance matrices have a limit.} The assumption 
 is plausible since the matrices are estimated from OLS residuals and most MR studies assume that the SNPs are independent of each other; {\color{black} see Section \ref{sec:corr_adjust} below and} Section 2.2. of \citet{zhao2020statistical} for discussions. Overall, Assumptions \ref{as:1}-\ref{as:3} are common in two-sample summary-data MR. Assumption \ref{as:4}, also known as weak IV asymptotics \citep{staiger65stock}, {\color{black} roughly states that the instrument's association to the exposure $\gamma_j$ has the same order of magnitude as the standard error of the estimated association, i.e. $1/\sqrt{n_2}$. Weak IV asymptotics has been widely used in econometrics as a theoretical device to better approximate finite-sample behavior of popular IV methods when instruments are weak. Also, weak IV asymptotics have informed useful guidelines for practice, most notably that practitioners should declare their instruments to be strong if the first-stage F statistic is greater than $10$ \citep{staiger65stock}.} Finally, except for Section \ref{sec:invalidAR}, we assume that the instruments are valid and focus most of the paper on weak instruments.

\subsection{Weak-IV Robust Tests for the Exposure Effect} \label{sec:Tests}
Consider the null hypothesis $H_0: \beta = \beta_0$ and the alternative $H_a: \beta \neq \beta_0$. We define two statistics $S(\beta_0) \in \reals^L$ and $R(\beta_0) \in \reals^L$ from the summary statistics $(\widehat{\Gamma}, \widehat{\Sigma}_{\Gamma}, \widehat{\gamma}, \widehat{\Sigma}_{\gamma})$. 
\begin{align} 
S(\beta_0) &= \left(\Sphi+ \beta_0^2 \Spi \right)^{-1/2} \left(\widehat{\Gamma}  - \beta_0 \widehat{\gamma} \right),  R(\beta_0) &= (\beta_0^2 \Sphi^{-1} + \Spi^{-1})^{-1/2} \left(\Sphi^{-1} \widehat{\Gamma}  \beta_0 +  \Spi^{-1} \hat{\gamma} \right). \nonumber
\end{align}
The statistics $S(\beta_0)$ and $R(\beta_0)$ are similar to the independent sufficient statistics of $\beta$ and \textcolor{black}{$\gamma$} in the traditional one-sample individual-data setting \citep{moreira2003conditional, moreira2019optimal} and in the two-sample individual-data setting  \citep{choi2018weak}. Notably, $S(\beta_0)$ and $R(\beta_0)$ can be computed with summary statistics and Lemma \ref{lemma1} shows that they maintain their independence properties under two-sample summary-data MR. 
\begin{lem} \label{lemma1}
If Assumptions \ref{as:1}-\ref{as:4} hold and $n_1/n_2 \to c \in (0,\infty)$,  $(S(\beta_0), R(\beta_0)) \inP (S_\infty(\beta_0), R_\infty(\beta_0))$ where 
$S_\infty \sim N[(\Sigma_1 + c \beta_0^2 \Sigma_2)^{-1/2} (\beta - \beta_0)C, I_L]$, 
$R_\infty \sim N[(\beta_0^2 \Sigma_1^{-1} + c^{-1}\Sigma_2^{-1})^{-1/2} ( \beta_0 \beta \Sigma_1^{-1} C + c^{-1} \Sigma_2^{-1} C), I_L]$, 
and $S_\infty \perp R_\infty$. Here, $I_L$ is an identity matrix of size $L$. 
\end{lem}
Using Lemma \ref{lemma1} along with a transformation technique in \citet{moreira2003conditional} and \citet{andrews2006optimal}, we can  
construct AR, K, and CLR tests for two-sample summary-data MR; we refer to them as mrAR, mrK, and mrCLR, respectively.
\begin{align}
T_{\rm mrAR}(\beta_0) &= Q_S(\beta_0), \quad{} T_{\rm mrK}(\beta_0) =  Q_{SR}^2(\beta_0) / Q_R(\beta_0) \nonumber \\
T_{\rm mrCLR}(\beta_0) &= \frac{1}{2} ( Q_S(\beta_0) - Q_R(\beta_0) +[ \left\{ Q_S(\beta_0) + Q_R(\beta_0) \right\}^2  - 4 \left\{Q_S(\beta_0) Q_R(\beta_0) - Q_{SR}^2(\beta_0) \right\}  ] ^\frac{1}{2} )  \nonumber
\end{align} 
Here, $Q_S(\beta_0) =S^\T (\beta_0)S(\beta_0)$,  $Q_{SR} =  S^\T (\beta_0) R(\beta_0)$, and $Q_R = R^\T (\beta_0) R(\beta_0)$. Suppose $\chi_k^2(1-\alpha)$ is the $1-\alpha$ quantile of a Chi-square distribution with $k$ degrees of freedom and ${\rm CDF}_{\chi_k^2}(x)$ is the cumulative distribution function of a Chi-square distribution with $k$ degrees of freedom. Theorem \ref{thm:tests} shows the asymptotic null distributions of $T_{\rm mrAR}(\beta_0), T_{\rm mrK}(\beta_0)$, and $T_{\rm mrCLR}(\beta_0)$.
\begin{theorem} \label{thm:tests}
Suppose Assumptions \ref{as:1}- \ref{as:4} and $H_0: \beta = \beta_0$ hold. For any $\alpha \in (0,1)$, 
as $n_1, n_2 \to \infty, n_1/n_2 \to c \in (0,\infty)$, we have 
\begin{align*}
&P\left(T_{\rm mrAR} > \chi_L^2(1-\alpha) \right) \rightarrow \alpha, \ P\left(T_{\rm mrK} > \chi_1^2(1-\alpha) \right) \rightarrow \alpha, \ P\left(w(T_{\rm mrCLR};Q_R) < \alpha \right) \rightarrow \alpha,
\end{align*}
where 
\[
w(x; y) = 1 - \frac{2G\left(\frac{L}{2} \right)}{ \sqrt{\pi} G\left( \frac{L-1}{2} \right) }\int\limits_{0}^{1} {\rm CDF}_{\chi_{L}^2}\left(\frac{x + y}{1 + y  \frac{z^2}{x}} \right)(1 - z^2)^{\frac{L - 3}{2}} dz
\]
and $G(\cdot)$ is the gamma function.
\end{theorem}
Theorem \ref{thm:tests} shows that under $H_0: \beta = \beta_0$, the two-sample summary data versions of the AR, K, and CLR tests converge to the same null distributions under the single-sample individual-data setting. In particular, like the original CLR test, 
$T_{\rm mrCLR}$ requires solving the integral $w(x;y)$ to obtain critical values; this integral can be computed by using off-the-shelf numerical integral solvers. Also, we can create asymptotically valid $1-\alpha$ confidence intervals for each test by ``inverting'' the test and retaining null $\beta_0$s that are accepted at level $\alpha$.
 {\color{black} Finally, we remark that if $\Sigma_1$ and $\Sigma_2$ are diagonal, the confidence interval from $T_{\rm mrAR}$ is almost identical to the confidence interval in Box 3 of  \citet{bowden2019improving} except our confidence interval uses $\chi_L^2(1-\alpha)$ whereas the latter uses $\chi_{L-1}^2(1-\alpha)$;  see Web Appendix B of the supporting information.}

{\color{black}
\subsection{Adjusting for Correlation from Marginal Regression Estimates} \label{sec:corr_adjust}
Often in two-sample summary-data MR, the summary statistics for each SNP are typically generated by running marginal regressions. If the SNPs are plausibly uncorrelated, which is often the case in practice through pre-processing, these summary statistics can be directly used in our tests. However, if the SNPs are correlated and the pairwise correlations are known a priori, say through the haplotype map (HapMap) or other genome-wide SNP annotation maps  \citep{johnson2008snap, burgess2016combining}, we provide a procedure to adjust the summary statistics so that our tests retain their asymptotic size control.

 
Formally, suppose we assume model \eqref{eq:model_iv}, but the summary-level statistics $\hat{\Gamma}, \hat{\gamma}$ and the diagonals of $\Sphi, \Spi$ are computed from simple linear regression with each IV. For each sample $l=1,2$, let $M_l \in \reals^{L \times L}$ be the pairwise correlation matrix between $L$ instruments. 
Because each GWAS may have different correlation structures, we allow for two, potentially different correlation matrices $M_l$. Theorem \ref{thm:corr} shows an equivalent result as Theorem \ref{thm:tests} when SNPs are correlated with a known $M_l$ and the summary statistics are based on simple linear regression estimates.
\begin{theorem} \label{thm:corr} Suppose the same assumptions in Theorem \ref{thm:tests} hold and for each sample $l=1,2$, let $M_l \in \reals^{L \times L}$ be the pairwise correlation matrix between $L$ instruments. Let $v_1,v_2, u_1, u_2 \in \reals^L$ where their $i$th elements equal to $v_{1i} = (\widehat{\Sigma}_{\Gamma,ii} n_1 + \widehat{\Gamma}_{i}^2)^{-1}$,  $v_{2i}  = (\widehat{\Sigma}_{\gamma,ii} n_2 + \widehat{\gamma}_{i}^2)^{-1}$, $u_{1i} = v_{1i} \widehat{\Gamma}_{i}$, and $u_{2i} = v_{2i} \widehat{\gamma}_i$. 
For each $l=1,2$, let $H_l \in \reals^{L \times L}$ where its $ij$th entry equals $M_{l,ij}\sqrt{v_{li} v_{lj}}$. Consider the following adjusted estimates, denoted as $\widetilde{\Gamma}, \widetilde{\gamma}, \widetilde{\Sigma}_{\gamma}$, and $\widetilde{\Sigma}_{\Gamma}$
\[
\widetilde{\Gamma} = H_1^{-1}u_1 , \quad{} \widetilde{\gamma} = H_2^{-1}u_2,\quad{} \widetilde{\Sigma}_{\Gamma} = \frac{1 - u_1^T H_1^{-1} u_1}{n_1 - L + 1}H_1^{-1}, \quad{}\widetilde{\Sigma}_{\gamma} = \frac{1 - u_2^T H_2^{-1} u_2}{n_2 - L + 1}H_2^{-1}.
\]
Then, mrAR, mrK, and mrCLR with $\widetilde{\Gamma}, \widetilde{\gamma}, \widetilde{\Sigma}_{\gamma}, \widetilde{\Sigma}_{\Gamma}$ have size control as defined in Theorem \ref{thm:tests}.
\end{theorem}
A limitation of Theorem \ref{thm:corr} is that the correlation matrices $M_{l}$ have to be known. To address this limitation, Section \ref{sec: correlatedIV} conducts a simulation study to study the impact of mis-specifying the correlation matrix $M_l$ on the performance of our tests.}
\subsection{Additional Properties of mrAR} \label{sec:add_mrAR}
This section discusses two brief extensions of mrAR motivated from the original AR test. First, we can take the minimum of $T_{\rm mrAR}$ to arrive at an MR-equivalent of the limited maximum likelihood estimator \citep{anderson1949estimation}; we refer to this estimator as mrLIML. 
\begin{equation} \label{eq:LIML}
\widehat{\beta}_{\rm mrLIML}:= \text{argmin}_{\beta_0} T_{\rm mrAR}(\beta_0)
\end{equation}
To compute $\widehat{\beta}_{\rm mrLIML}$, we can (i) do a simple grid search or (ii) gradient descent. 
{\color{black} Theorem \ref{thm:mrLIML} shows that if $\Sphi$ and $\Spi$ are diagonal, $\widehat{\beta}_{\rm mrLIML}$ is equivalent to estimators proposed by \citet{zhao2020statistical} and  \citet{bowden2019improving} and thus, $\widehat{\beta}_{\rm mrLIML}$ can be thought of as a generalization of these two estimators under correlated instruments.
\begin{theorem} \label{thm:mrLIML} If $\Sphi$ and $\Spi$ are diagonal, $\widehat{\beta}_{\rm mrLIML}$ is equivalent to the estimator in equation (3.2) of \citet{zhao2020statistical} and the estimator in Box 2 of \citet{bowden2019improving}
\end{theorem}}
With Theorem \ref{thm:mrLIML}, mrLIML inherits the statistical properties described in Theorems 3.1 and 3.2 of \citet{zhao2020statistical}, notably consistency and asymptotic Normality. Also, Theorem \ref{thm:mrLIML} illustrates that \citet{zhao2020statistical}'s estimator can be motivated based on a testing framework. {\color{black} Finally, \citet{windmeijer2019two} showed a variant of $\widehat{\beta}_{\rm mrLIML}$ is a minimum-distance estimator.}


 Second, like the original AR test, mrAR can be indirectly used to detect presence of invalid instruments by checking whether the confidence interval is empty \citep{davidson2014confidence}; see Section \ref{sec:invalidAR} for numerical illustrations. 
{\color{black} This property of mrAR is similar to the test for pleiotropy using the Q statistic in \citet{bowden2017framework} and \citet{bowden2019improving}. However, the Q statistic is testing the null hypothesis that the exposure effects derived from each SNP are equal to each other and rejecting this null indicates that the exposure effects are heterogeneous. In contrast, mrAR is testing that the true exposure effect is $\beta_0$ and rejecting this null indicates that $\beta_0$ is not the true effect or that the model assumptions under the null are not plausible. Given the slight differences in the null hypotheses, the Q statistic has $\chi_{L-1}^2(1-\alpha)$ as its null distribution whereas mrAR has $\chi_L^2(1-\alpha)$ as its null distribution. Also, the Q statistic requires independent instruments whereas mrAR can handle correlated instruments. But, standard pre-processing of data in two-sample summary-data MR (see Section 4.1 for one example) can reduce concerns for correlated instruments in the Q statistic.}

\section{Simulations} \label{sec:simu}
We conduct simulation studies to study the performance of our test statistics. The data is generated from the structural model in \eqref{eq:model_iv} with $n_1 = n_2 = 100,000$, which are on the same order as the sample size from our data analysis in Section \ref{sec:data}. The random errors $(\delta_{1i},\delta_{2i})$ are generated from a bivariate standard Normal and the random error $\epsilon_{li}$ is equal to $\epsilon_{li} = \rho \delta_{li} + (1- \rho^2)^{1/2} e_{li}$; the term $e_{li}$ is from an independent standard Normal and $\rho = 0.1$. We remark that $\rho$ reflects the endogeneity between the outcome and the exposure. The $L = 100$ instruments take on values $0,1,2$, similar to how SNPs are recorded in GWAS{\color{black}. Except for Section \ref{sec: correlatedIV},} the instruments are generated independently from a Binomial distribution Binom$(2,p_j), j = 1,\cdots,L$ with $p_j$ drawn from a uniform distribution Unif$(0.1,0.9)$. After generating individual-level data, we compute the summary statistics for sample $l=1$, i.e. $\widehat{\Gamma}$ and $\widehat{\Sigma}_{\Gamma}$, by running an OLS regression between $Y_{1i}$ and $Z_{1ij}$ for each instrument $j$ and extracting the estimated coefficient and its standard error. Similarly, we compute the summary statistics for sample $ l=2$, i.e. $\widehat{\gamma}$ and $\widehat{\Sigma}_{\gamma}$, by running an OLS regression between $D_{2i}$ and $Z_{2ij}$ for each instrument. The simulation varies the exposure effect $\beta$ and the IV-exposure relationship $\gamma$. Specifically, the value of $\gamma$ ranges from $\{ (r-0.5)/n_1\}^{1/2}$ to $\{ (r + 0.5)/n_1\}^{1/2}$ to mimic weak-IV settings as formalized in Assumption \ref{as:4} and {\color{black} we let $r$ be $1$, $4$, $16$, and $25$}. We remark that $r$ roughly corresponds to the first-stage $F$ statistic common in IV strength testing; {\color{black} see Section \ref{sec:data_1} below and Web Appendices A and B of the supporting information.}
The simulation is repeated $1,000$ times.

 For comparison, we use the following methods in two-sample summary-data MR: {\color{black} robust MR-Egger regression via a radial regression with second-order correction \citep{bowden2018improving} (MR-Egger.r)}, the profile likelihood estimator with a squared error loss \citep{zhao2020statistical} (MR-RAPS), and the weighted median estimator \citep{bowden2016consistent} (W.Median). {\color{black} As mentioned in Section \ref{sec:add_mrAR}, the exact IVW estimator in Box 2 of \citet{bowden2019improving} is equivalent to MR-RAPS with a squared error loss}. We use the software \verb+Mendelianrandomization+ \citep{yavorska2017mendelianrandomization},  {\color{black}\verb+RadialMR+ \citep{bowden2019improving} }, 
and \verb+mr.raps+ \citep{zhao2020statistical} to run these methods.

\subsection{Size and Power of Proposed Tests} \label{sec:size_power}
\begin{figure}
\begin{center}
\includegraphics[width=\textwidth]{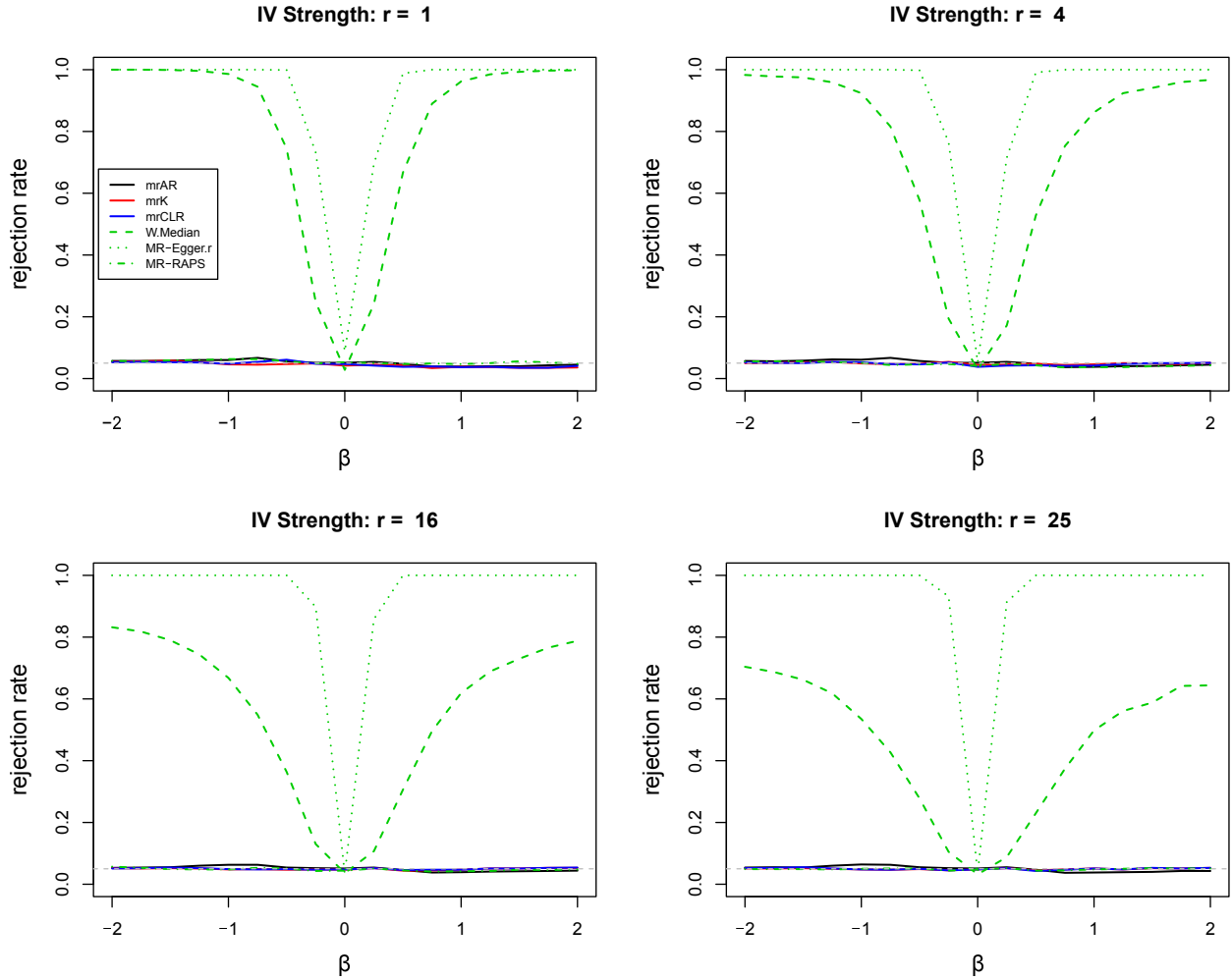}
\caption{Size under different IV strength. The number of instruments is $L=100$. $r$ represents instrument strength, with higher indicating stronger instruments; $r$ approximately corresponds to the first-stage F statistic for IV strength. The grey dashed horizontal line represents the Type I error rate of $0.05$. This figure appears in color in the electronic version of this article, and any mention of color refers to that version.}
\label{size}
\end{center}
\end{figure}

First, we examine the size of the proposed tests. Figure~\ref{size} shows the Type I error under $H_0: \beta = \beta_0$ with $\beta_0$ ranging from $-2$ to $2$. All the tests are carried out under the null. {\color{black} Except for MR-RAPS, at each value of $r$, the size distortion from pre-existing methods increases as the true exposure effect $\beta$ moves away from the null effect of $\beta_0 = 0$. In contrast, MR-RAPS and our tests have size control for every value of $r$, validating Theorem \ref{thm:tests}. }

\begin{figure}
\centerline{\includegraphics[width = \textwidth]{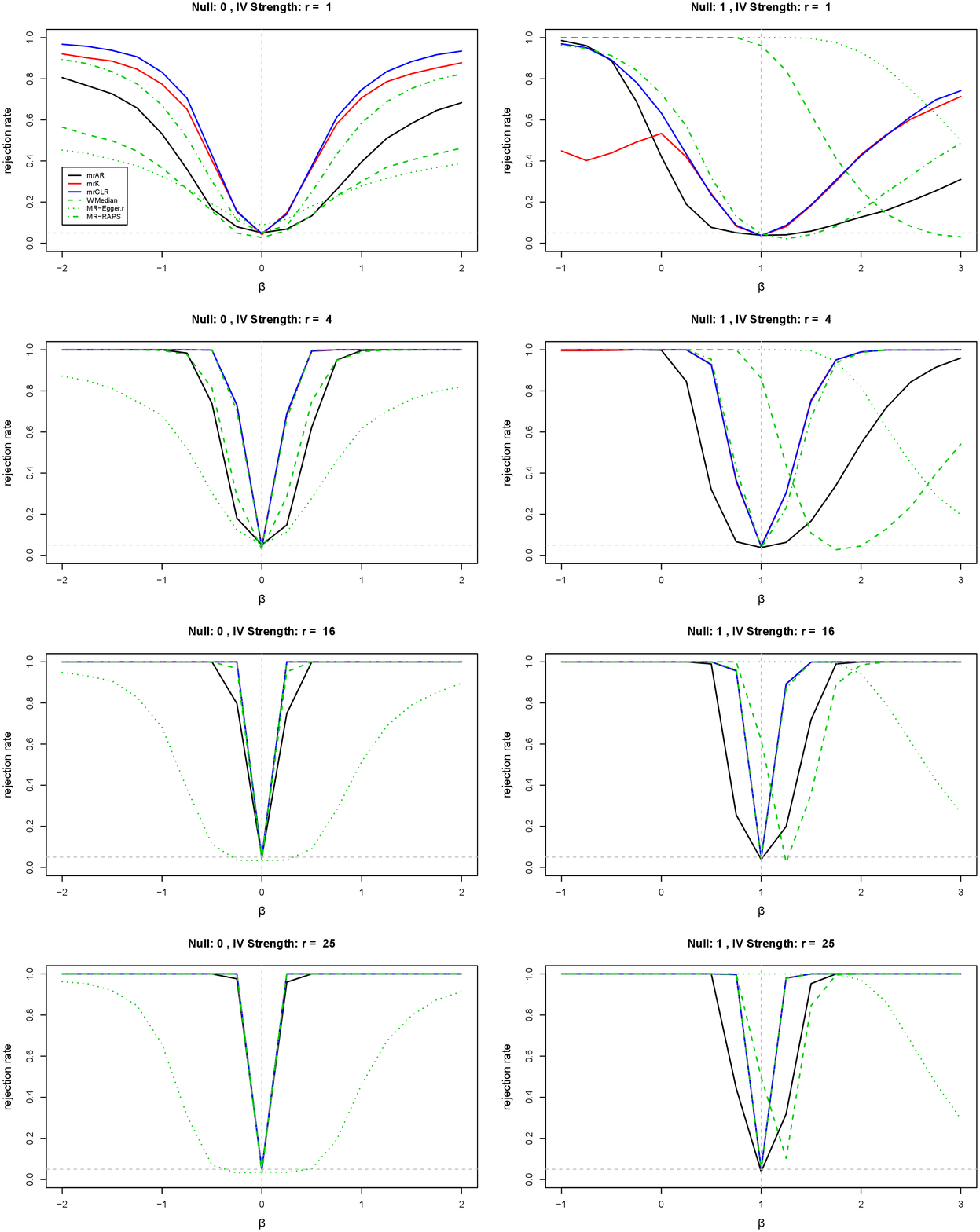}}
\caption{Power curves under different nulls and IV strength. The left panel is under $H_0: \beta = 0$ and the right panel is under $H_0: \beta = 1$; each null value is represented by a grey dashed vertical line. $r$ represents instrument strength, with higher indicating stronger instruments; $r$ approximately corresponds to the first-stage F statistic for IV strength. The grey dashed horizontal line represents the Type I error rate of $0.05$. This figure appears in color in the electronic version of this article, and any mention of color refers to that version.}
\label{tests}
\end{figure}

Next,  
Figure~\ref{tests} examines statistical power when the null hypothesis is $H_0: \beta = 0$ (left panel) or $H_0: \beta = 1$ (right panel); the significance level is set at $\alpha = 0.05$. Under the null hypothesis of no effect $H_0: \beta = 0$, { all methods correctly control the Type I error for every value of $r$ and \textcolor{black}{every method except MR-Egger.r has similar statistical power when $r =16$ or $r=25$}. However, for lower values of $r$, mrK and mrCLR have superior statistical power, with mrCLR having the best power among all methods; this agrees with \citet{andrews2006optimal} who showed that the CLR test in the single-sample individual-data setting is nearly optimal. Under $H_0: \beta = 1$, none of the pre-existing methods except MR-RAPS have Type I error control for every value of $r$. In contrast, our tests always maintain Type I error control and \textcolor{black}{mrCLR has the best power among all existing methods presented in the simulation study}. Section 1.1 of the supplementary materials repeats the simulation above with $r = 50$ and the results are similar to the case when $r=25$. }

{\color{black} Overall, our tests always retain size control for different values of $r$ and null values $\beta_0$. Additionally, mrCLR has the best power among existing methods when $r$ is low and has no worse power than existing methods when $r$ is high. In short, mrCLR does well under both weak and strong instruments. }

\subsection{mrAR Under Invalid Instruments} \label{sec:invalidAR}


In this section, we explore mrAR's ability to detect a wide variety of invalid instruments under different $r$. The simulation model is generated from the structural model:
\begin{align} \label{eq:model_invalid}
\begin{split}
&\ Y_{li} = \beta_{\rm int} + D_{li} \beta + Z_{li}^\intercal \alpha + \epsilon_{li},\quad{}  D_{li} = \gamma_{\rm int} + Z_{li}^\intercal \gamma+ \delta_{li}, \quad{} E[\epsilon_{li}, \delta_{li} \mid Z_{li}] = 0.
\end{split}
\end{align}
Compared to model \eqref{eq:model_iv}, model \eqref{eq:model_invalid} includes $\alpha$, which is the direct effect of the instruments $Z_{li}$ on the outcome $Y_{li}$. If $\alpha = 0$, there is no direct effect and all instruments are valid. If $\alpha \neq 0$, instruments with $\alpha_j \neq 0$ have direct effects on the outcome and are invalid; see \citet{kang2016instrumental} and \citet{bowden2016assessing} for additional discussions of defining invalid IVs.

For the simulation study, we vary the number of invalid instruments, i.e. $\sum_{j=1}^{L} I(\alpha_j \neq 0)$, from 10\% of $L$ to 90\% of $L$. We also vary the magnitude of $\alpha_j$ to be either $0.003$, $0.01$, or $0.05$;  $0.003$ roughly corresponds to $\alpha_j \approx 1/\sqrt{n_1}$, or essentially local-to-zero invalid instruments, $0.05$ roughly corresponds to $\alpha_j \approx 1/n_1^{1/4}$, or non-local invalid instruments, and $0.01$ is somewhere in between these types of invalid instrument. \textcolor{black}{The invalid instruments do not satisfy the Instrument Strength Independent of Direct Effect (InSIDE) assumption \citep{bowden2015mendelian}. 
Also, when less than 50\% of the instruments are invalid, the majority and the plurality assumptions \citep{guo2018confidence} are satisfied. When 50\% or more of the instruments are invalid, they are not satisfied.} The true causal effect $\beta$ is set at $0.5$. and we test $H_0: \beta = \beta_0$ against $H_1: \beta \neq \beta_0$ where $\beta_0$ ranges from $-4$ to $5$. The other parameters of the simulation settings remain the same as before. {\color{black} Finally, for comparison, we use the modified second-order weighting Q statistic in \citet{bowden2019improving}, which was designed to test for invalid instruments.}

\begin{figure}
\centerline{\includegraphics[width = \textwidth]{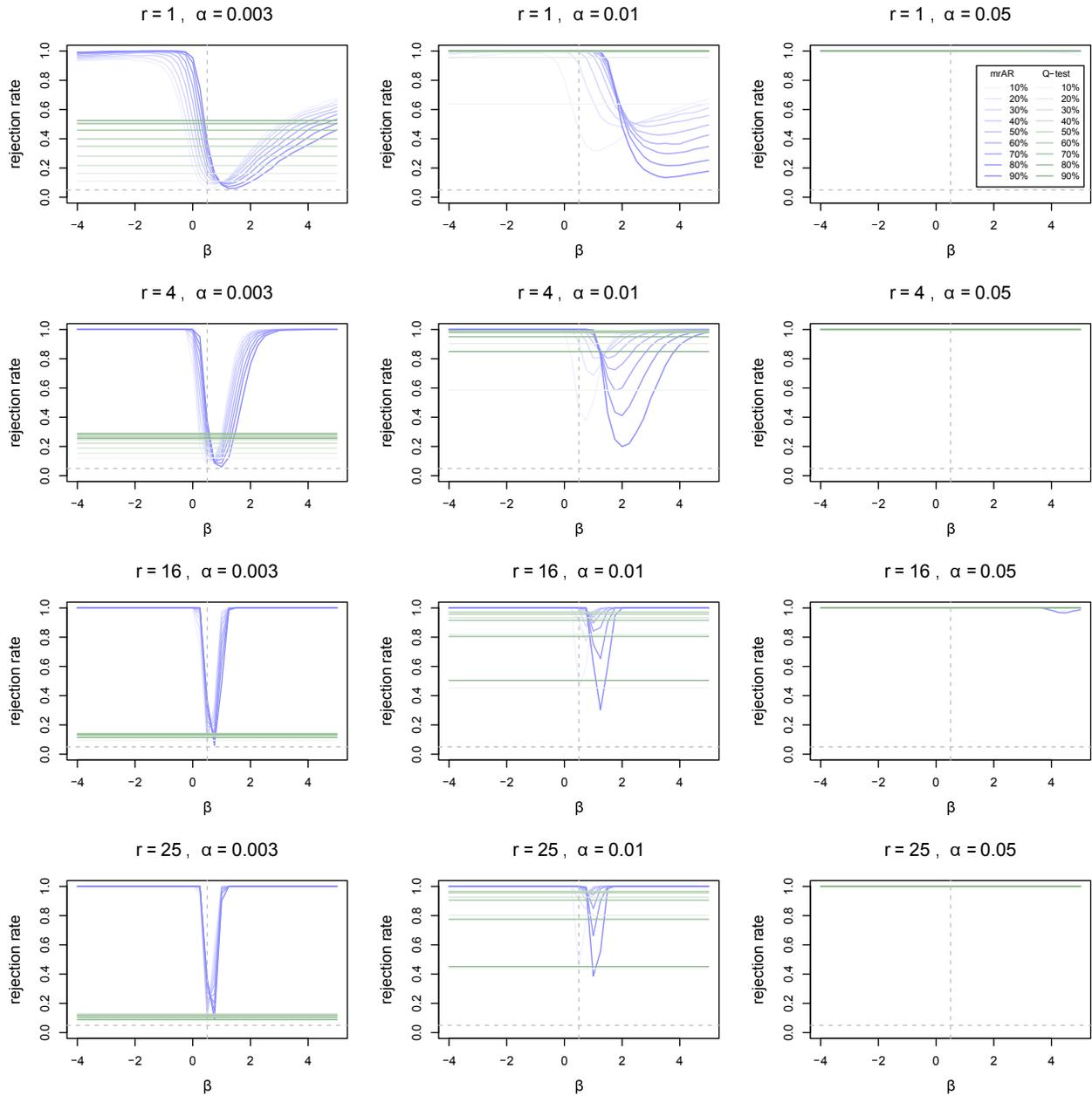}}
\caption{Rejection rates of mrAR and Q-test under invalid instruments. The true value of $\beta$ is represented by the vertical grey line at $0.5$ and we test $H_0: \beta = \beta_0$ for different values of $\beta_0$. $r$ represents instrument strength, with higher indicating stronger instruments; $r$ approximately corresponds to the first-stage F statistic for IV strength. Each column changes the magnitude of $\alpha$ from $0.003$ to $0.05$. Each purple or green line represents the proportion of invalid IVs as measured by the number of $\alpha_j \neq 0$ divided by $L$. The grey horizontal line represents the Type I error rate of $0.05$. This figure appears in color in the electronic version of this article, and any mention of color refers to that version.}
\label{invalid}
\end{figure}

{\color{black}
Figure \ref{invalid} shows the rejection rates of mrAR and the Q statistic at each value of $\beta_0$ under different configurations of $\alpha$. Note that because the Q statistic tests the null hypothesis of heterogeneous effect and doesn't rely on a null value $\beta_0$ for the true effect, it shows a flat rejection rate across $\beta_0$.}
For different types of $\alpha$ and instrument strength $r$, mrAR remains above the $0.05$ threshold across all $\beta_0$, thereby always rejecting the null hypothesis and returning empty confidence intervals to alert investigators about potentially invalid instruments. The mrAR test rejects more frequently when $\alpha$ is farther away from zero or when instrument strength increases. For example, if $\alpha_j = 0.05$, which is roughly on the order of $1/n_1^{1/4}$, mrAR always rejects at 100\% regardless of the proportion of invalid instruments. {\color{black} The result of the Q statistic is similar to the mrAR as the Q statistic remains above the $0.05$ threshold for all values of $r$. Also, similar to mrAR, the Q statistic has a higher rejection rate when the magnitude of $\alpha_j$ grows. Web Appendix A of the supporting information repeats the simulation above with $r = 50$ and another test statistic by \citet{bowden2019improving} and the results are similar to the case when $r=25$. Overall, we believe both mrAR and the Q statistic are useful in detecting invalid instruments under a wide variety of settings and could be used as pre-tests for methods that are sensitive to invalid instruments.} 

{\color{black} 
\subsection{Correlated Instruments} \label{sec: correlatedIV}


In this section, we assess our proposed tests under correlated instruments, especially demonstrating our adjustment procedure in Section \ref{sec:corr_adjust} and the procedure's performance under incorrectly specified correlation matrices $M_l$. Under model \eqref{eq:model_iv} and for each $l=1,2$, we set the diagonal elements of $M_l$ as $1$ and the non-diagonal, $ij$th element of $M_l$ as $\rho = 0.3$ if $\lvert i - j\rvert = 1$, and $0$ if $\lvert i - j\rvert \geq 2$; in short,
adjacent SNPs are correlated with correlation coefficient $\rho = 0.3$. For each $l$, we let $\widehat{M}_l$ denote our working value of $M_l$ and use it in our adjustment procedure. If $\widehat{M}_{l} = M_l$, we say that the correlation matrix $\widehat{M}_{l}$ is correctly specified and if $\widehat{M}_{l} \neq M_l$, the correlation matrix is incorrectly specified. 

For incorrect specifications of $\widehat{M}_{l}$, we assume the zero elements of $\widehat{M}_{l}$ are the same as $M_l$, but the non-zero, off-diagonal elements differ, taking on values $\hat{\rho} \in \{0, 0.2, 0.3, 0.4\}$. When $\hat{\rho} < \rho = 0.3$, the investigator is assuming there is little to no correlation between instruments, even though in reality, instruments are actually correlated. When $\hat{\rho} > \rho = 0.3$, the investigator is assuming there is a stronger correlation than the true correlation of $0.3$. When $\hat{\rho} = \rho = 0.3$, the investigator has correctly specified the true correlation. The other parameters are the same as in Section \ref{sec:size_power} except we only look at $r = 1$ to assess our proposed tests in the hardest scenario for instrument strength; Section 1.3 of the supplementary materials contain cases when $r$ is higher and shows similar results.

Figure \ref{corr} shows the size and power of our tests based on the adjustment procedure described in Section \ref{sec:corr_adjust}. We see that if $\hat{\rho} \leq \rho$, our adjusted tests have a slight size inflation not exceeding 10\% across different values of $\beta_0$. Also, the size inflation decreases as $\hat{\rho}$ gets close to $\rho$. For power, mrCLR has the best power among the three tests. When $\hat{\rho} = \rho$, our adjusted tests have size control at level $\alpha$, as predicted by Theorem \ref{thm:corr}, and exhibit the same behavior demonstrated in Section \ref{sec:size_power}. However, when $\hat{\rho} > \rho$, the size distortion of our adjusted tests ranges from $1\%$ to $25\%$, with mrK and mrCLR having the smallest size distortions. 

\begin{figure}
\centerline{\includegraphics[width = \textwidth]{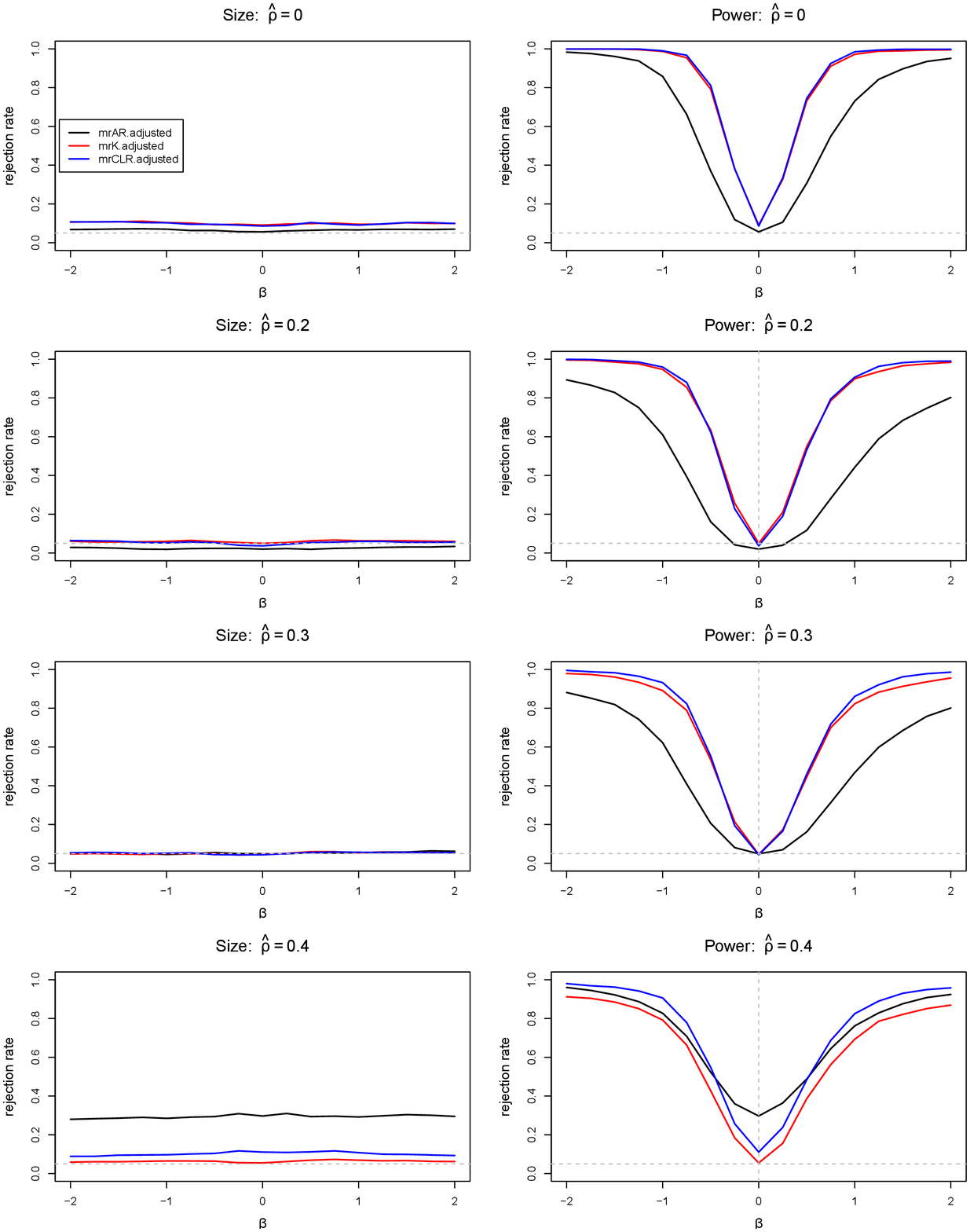}}
\caption{Performance of proposed tests with correlated instruments when $r=1$. The left panel shows size and the right panel shows power of testing $H_0: \beta = 0$. The true correlation $\rho$ is $0.3$ and each row changes the magnitude of $\hat{\rho}$ from $0$ to $0.4$. The true value of $\beta$ in the power plot is represented by a grey dashed vertical line at $0$. The grey dashed horizontal line represents the Type I error rate of $0.05$. This figure appears in color in the electronic version of this article, and any mention of color refers to that version.}
\label{corr}
\end{figure}

Web Appendix A of the supporting information explores other mis-specifications of the correlation matrix, notably the case where we mis-specify the zero elements. Overall, while the simulation studies do not capture every possible mis-specification, the main takeaway message is that if an investigator has access to reasonable estimates of the correlation matrices, he/she should use the proposed adjustment procedure for our tests. However, if there is uncertainty about the correlation matrices, he/she should proceed as if there is no correlation between the instruments, even though in reality they are actually correlated. In this case, the sizes of our $0.05$ level tests are modestly inflated, with sizes ranging from 1\% to 10\%. In contrast, if one adjusts our tests for correlation even though there isn't any correlation between instruments, the size distortions range from 1\% to 25\%.}

\section{Data Analysis}  \label{sec:data}
\subsection{Replication of \citet{zhao2020statistical}'s Empirical Study} \label{sec:data_1}
To validate our methods in real data, we replicate an analysis by \citet{zhao2020statistical} on the effect of BMI on systolic blood pressure where the effect was known to be positive. The authors used three independent GWAS, one from the UK Biobank GWAS (SBP-UKBB) and the other two from GWAS by the Genetic Investigation of ANthropometric Traits (GIANT) consortium \citep{locke2015genetic}. Specifically, the ``BMI-MAL'' and ``SBP-UKBB'' datasets in \citet{zhao2020statistical} provided summary statistics of the IV-exposure and IV-outcome statistics, respectively. The ``BMI-FEM''  dataset from \citet{zhao2020statistical} was used to pre-screen for strong and uncorrelated IVs. {\color{black} This pre-screening dataset led to two sets of instruments, a set of 160 instruments based on the p-value threshold of $10^{-4}$ and another set of 25 instruments based on the genome-wide significance level of $ 5 \cdot 10^{-8}$; the latter is also the default setting for selecting instruments in the MR-Base platform \citep{hemani2018mr}. In both sets of instruments, each pair of instruments were at least 10,000 kilobases apart and had linkage disequilibrium correlation coefficients less than $0.001$; see \citet{zhao2020statistical} for more details.}. 

We compute 95\% confidence intervals using mrAR, mrK, mrCLR, W.Median, MR-Egger.r, and two versions of MR-RAPS, MR-RAPS with a square error loss (MR-RAPS) and MR-RAPS with a huber loss (MR-RAPS.r). \textcolor{black}{We also computed the F statistic typical in IV studies to measure instrument strength. Specifically, in Web Appendix A of the supporting materials, we show that if the instruments are independent of each other, the usual F statistic in linear IV models can be computed from two-sample summary-data MR by using the formula
\[
F = \left( \frac{n_2 - L + 1}{L} \right)\left( \frac{ \sum_{i=1}^{L} \frac{F_i}{F_i + n_2 - L-1}}{1- \sum_{i=1}^{L} \frac{F_i}{F_i + n_2 - L-1}} \right)
\]
 where $F_i = \widehat{\gamma}_{i}^2/\Sigma_{\gamma,ii}$ is the square of the t-statistic for instrument $i$, or equivalently the F statistic for instrument $i$. If $n_2 - L$ is sufficiently large enough, we can approximate the $F$ statistic as the average of $F_i$s, i.e. $F \approx L^{-1} \sum_{i=1}^{L} F_i$; this latter approximation has been mentioned in prior works \citep{bowden2019improving} and our exposition in the supplementary materials provide a formal justification to this approximation.} 
 
\begin{table}
\caption{95\% confidence intervals of the effect of body mass index on systolic blood pressure from \citet{zhao2020statistical}. The thresholds refer to significance thresholds used in \citet{zhao2020statistical}'s analysis to select instruments.}
\begin{center}
\begin{tabular}{l ccc}
\hline
        & 25 Instruments ($5\times 10^{-8}$ Threshold) & 160 Instruments ($10^{-4}$ Threshold) \\ \hline
 ${\rm mrK}$    &   (0.205, 0.530)    & (0.377, 0.771) \\
 ${\rm mrCLR}$   & (0.211, 0.524)  &  (0.415, 0.731)  \\
${\rm mrAR}$  &  ($\emptyset$) & ($\emptyset$) \\
 W.Median  & (0.278, 0.762) &  (0.318, 0.726)\\
 MR-Egger.r  & (0.075, 1.038)   & (0.203, 0.573)   \\
 MR-RAPS  &    (0.221, 0.514) & (0.499, 0.712)   \\ 
 MR-RAPS.r  &   (0.097, 0.610)  &  (0.141, 0.615) \\
 F statistic & 58.140 &  25.462 \\
 Q statistic (p-value) &5.582 $\times 10^{-8}$ &  5.727$ \times 10^{-61}$ \\
 \hline
\end{tabular}
\label{BMI-SBP}
\end{center}
\end{table}
 Table \ref{BMI-SBP} summarizes our results and two interesting observations emerge from the re-analysis of  \citet{zhao2020statistical}'s dataset. First, almost all methods have overlapping 95\% confidence intervals, despite both mrAR and the Q statistic alerting investigators that invalid instruments may be present. Specifically, mrK and mrCLR, which are not robust to invalid instruments, generate similar confidence intervals as methods that are robust to invalid instruments, notably W.Median, MR-Egger.r, and MR-RAPS.r. This suggests either that invalid instruments are present, but small or, as \citet{small2008war} suggests, when IVs are weak, the first-order bias is dominated by weak IVs and therefore, correcting for weak-IV bias through weak-IV robust methods like ours can attenuate biases from invalid IVs. 
 Note that Table \ref{BMI-SBP} only \textcolor{black}{reports} the positive region since we know a priori that the effect is positive; in general, when the exposure effect direction is unknown, we recommend taking the union of the disjoint intervals.

\begin{figure}
\centerline{\includegraphics[width = \textwidth]{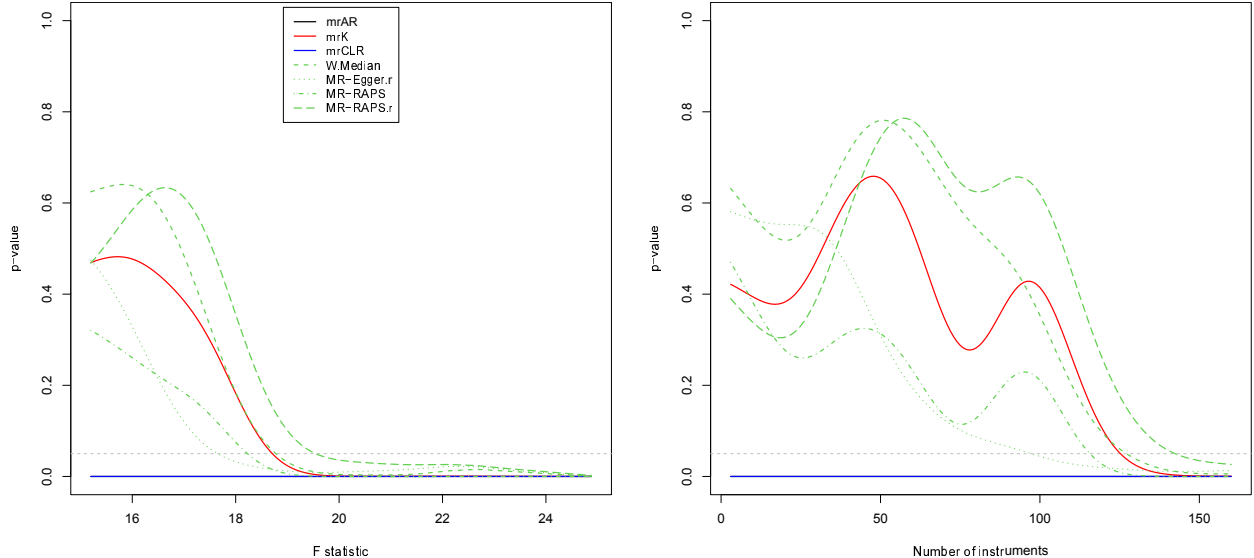}}
\cprotect\caption{P-values for testing the null hypothesis of no exposure effect under different number of instruments and strength. We plot a smoothed curve of p-values using a Gaussian kernel with the p-values' standard deviation as the smoothing parameter; this is the default value in the \verb+sm.regression()+ function in the R package \verb+sm+. The grey dashed horizontal line represents the rejection threshold at $0.05$. The number of instruments goes from $3$ weakest instruments to all $160$ instruments. This figure appears in color in the electronic version of this article, and any mention of color refers to that version.}
\label{pval_Fstat}
\end{figure}

{\color{black}
\subsection{Validation: Testing the Null of No Effect} \label{sec:data_2}
Given that the effect of BMI on SBP is generally thought to be positive, we validate our analysis above by testing the null hypothesis of no exposure effect (i.e $H_0: \beta = 0$), but with an increasing set of strong instruments. Testing the null of no effect is one of the first questions that an MR investigator may ask about the study and for the BMI study, an ideal test should reject the null regardless of the strength of the instruments. In other words, when we go from the weakest set of instruments to all 160 instruments, an ideal test should reject the null of no effect by generating a p-value that is less than $\alpha = 0.05$. We remark that this validation analysis is similar in spirit to the power simulations in Section \ref{sec:size_power} where we are examining the power to reject the null of no effect in favor of a known true alternative. 

Figure \ref{pval_Fstat} plots the p-values from testing this null hypothesis as a function of the F statistic and the number of instruments, arranged in increasing order of strength. We see that mrCLR always \textcolor{black}{rejects} the null hypothesis of no effect of BMI on blood pressure at $\alpha = 0.05$ level, even with the 3 weakest IVs (left end of the x-axis), while other MR methods cannot reject the null hypothesis unless more strong IVs are present. mrAR and mrK also have smaller p-values for testing the null than existing methods across all instrument strength. These observations also confirm our simulation study in Section \ref{sec:size_power} where mrCLR exhibited superior power compared to existing methods. However, as a practical matter, we caution that relying solely on p-values to establish a causal effect can potentially be misleading since they don't tell the magnitude of the true effect nor reveal the direction of potential biases.} 

\subsection{Validation: Stress Testing Based on \citet{bound1995problems}} \label{sec: stressTesting}
To further validate our analysis, we follow an approach by \citet{bound1995problems} where we ``stress-test'' methods by replacing each of the original IV-exposure \textcolor{black}{effects} $\hpi_j$ and the IV-outcome \textcolor{black}{effects} $\widehat{\Gamma}_j$ for the set of 160 instruments by $\widehat{\gamma}_{j}^{\rm new} \sim N(K \hpi_j, \widehat{\Sigma}_{\gamma,j})$ and $\widehat{\Gamma}_{j}^{\rm new} \sim N(K \hpi_j \beta, \widehat{\Sigma}_{\Gamma,j})$; here, $\beta$, the true exposure effect, is set to be $0.5$ and $1.5$. The parameter $K$ controls IV strength and ranges from $0$ to $1$. 
Under $K=1$, the new IV-exposure and IV-outcome effects are essentially the original effects, but with a known true value of $\beta$. But, as $K$ decreases to $0$, the IV becomes weaker than the original ones. In the extreme case when $K = 0$, there is no way to consistently estimate $\beta$; the new IV-exposure and IV-outcome effects look statistically indistinguishable if the true exposure effect is $\beta = 1$ or $\beta = 1000$. 

\begin{figure}
\centerline{\includegraphics[width=6in]{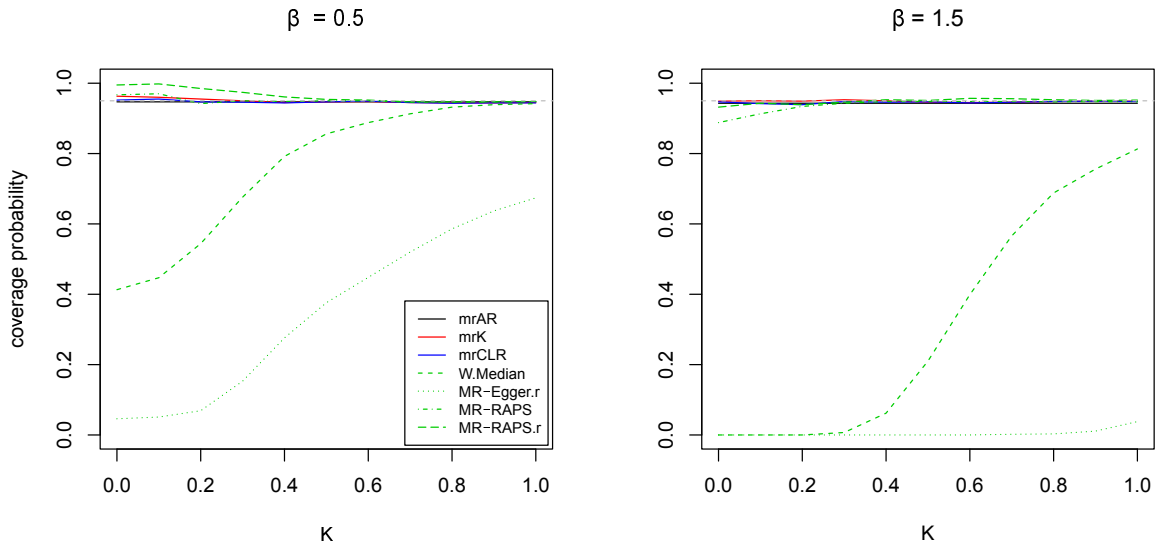}}
\caption{Coverage probability under different IV strength. The left panel sets the true causal effect $\beta$ to be 0.5 and the right panel sets $\beta$ to be 1.5. The grey dashed horizontal line represents $95\%$ coverage. This figure appears in color in the electronic version of this article, and any mention of color refers to that version.}
\label{coverage}
\end{figure}

An ideal confidence interval should be able to (i) automatically detect the lack of identification of the exposure effect $\beta$ by producing an infinite confidence interval when $K = 0$ \textcolor{black}{95\% of the time} to maintain a 95\% coverage rate and (ii) a bounded confidence interval with a 95\% coverage rate when $K$ moves away from zero; in short, it should produce 95\% coverage rates across all values of $K$. As Figure~\ref{coverage} shows, the existing MR methods do not achieve these two goals. For example, when $K=0$ and $\beta = 0.5$, W.Median, MR-Egger.r, MR-RAPS, and MR-RAPS.r produced bounded intervals across 1,000 simulations even though $\beta$ is not identifiable. We remark that it is by luck that MR-RAPS' confidence intervals produced large, bounded intervals to cover the true effect of $\beta$ of $0.5$ even though $\beta$ is not identifiable in this scenario. Indeed, in the \textcolor{black}{right-hand} plot of  Figure~\ref{coverage} when $K = 0$ and $\beta = 1.5$, W.Median, MR-Egger.r, MR-RAPS, and MR-RAPS.r again produced bounded intervals, but their coverage rates drop, sometimes dramatically, because the true effect is farther away from $0$. In contrast, mrAR, mrK, and mrCLR produced infinite confidence intervals when $K = 0$ \textcolor{black}{around 95\% of the time to maintain 95\% coverage}. More generally, our tests always maintained a 95\% coverage rate across all values of $K$ and $\beta$ and satisfied the two \textcolor{black}{criteria} (i) and (ii).

\section{Summary, Limitations, and Recommendations}
In this paper, we propose weak-IV robust test statistics for two-sample summary data in MR by extending the existing AR, Kleibergen, and CLR tests in econometrics and show that they have Type I error control under weak instrument asymptotics.  {\color{black} The simulation results show that if there is no evidence of pleiotropy, then among existing tests designed for weak instruments, mrCLR is superior.} 
Similarly, the replication of the data analysis in \citet{zhao2020statistical} and the two validation analyses echoed findings from the simulation studies. 


{\color{black} While this work is focused on weak instruments in MR and using powerful tests from econometrics to address them, a major concern in MR is the presence of invalid instruments. Invalid instruments would violate Assumption (A2) and can bias our confidence intervals. However, as we saw in the replication analysis in Section \ref{sec:data_1}, the intervals from our methods, which are not robust to invalid instruments, were similar to those from methods that are robust to invalid instruments, suggesting that invalid instruments had little impact in this particular study or, as mentioned earlier, correcting for weak-IV bias through weak-IV robust methods like ours can attenuate biases from invalid IVs. 

Web Appendix A of the supporting information conducts a simulation study to assess the sensitivity of our methods under various types of invalid instruments. In summary, the results show that mrK and mrCLR are reasonably robust against invalid instruments when the direct effects are small. But, they are significantly biased as the magnitude of the direct effects and the number of invalid instruments increase; we also show that other robust methods, notably MR-RAPS.r, suffer in this setting as well. 
Thankfully, as Section \ref{sec:invalidAR} showed, mrAR is able to detect invalid instruments in a variety of settings and can be used as a pre-screening test before using our methods; see \citet{kang2020two} for details. }

{\color{black} We conclude by making some recommendations about how to use our methods in practice. First, from the simulation results and the data analysis above, we recommend practitioners start by using mrAR or the Q statistic to check for invalid instruments, especially if the instruments are weak. Second, if mrAR returns non-empty confidence intervals and thus, the instruments are plausibly valid, we recommend investigators use mrCLR. Additionally, mrCLR is the only test that we know in the two-sample summary-data MR which satisfies \citet{dufour1997some}' necessary condition for valid confidence intervals and adapts to produce infinite confidence intervals, if necessary. Third, if mrAR returns empty confidence intervals, suggesting a presence of invalid instruments, there is currently no method that can be robust to every parametrization of invalid instruments. As such, we recommend practitioners follow what we did in our empirical analysis and 
use a combination of methods. In particular, as we saw in the empirical example above, even though mrAR returned an empty confidence interval and suggested invalid instruments were present, the confidence intervals generated from mrCLR and mrK were similar to those from methods that are robust to invalid instruments, say MR-RAPS with a robust loss function, and comparing the confidence intervals between these methods yielded some understanding about the type of invalid instruments present in the study. Fourth, from our investigation into correlated instruments, when using our tests, if the pairwise correlation between instruments is unknown, we generally recommend investigators \textcolor{black}{proceed as if} instruments aren't correlated, even if in reality they are actually correlated,  as this strategy tends to lead to only modest size distortions.}

\section*{Acknowledgements}
The research of Hyunseung Kang was supported in part by NSF Grant DMS-1811414. The research of Sheng Wang was supported in part by the University of Wisconsin-Madison's Data Science Initiative Grant.

\section*{Data Availability Statement}

The data that support the findings of this paper are openly available in CRAN at https://cran.r-project.org/package=mr.raps, reference number \citep{mrrapsSoftware}.

\bibliography{paper-ref}

\begin{thebibliography}{}

\bibitem[Anderson et~al., 1949]{anderson1949estimation}
Anderson, T.~W., Rubin, H., et~al. (1949).
\newblock {Estimation of the parameters of a single equation in a complete
  system of stochastic equations}.
\newblock {\em {The Annals of Mathematical Statistics}}, 20(1):46--63.

\bibitem[Andrews et~al., 2006]{andrews2006optimal}
Andrews, D.~W., Moreira, M.~J., and Stock, J.~H. (2006).
\newblock {Optimal two-sided invariant similar tests for instrumental variables
  regression}.
\newblock {\em {Econometrica}}, 74(3):715--752.

\bibitem[Bound et~al., 1995]{bound1995problems}
Bound, J., Jaeger, D.~A., and Baker, R.~M. (1995).
\newblock {Problems with instrumental variables estimation when the correlation
  between the instruments and the endogenous explanatory variable is weak}.
\newblock {\em {Journal of the American Statistical Association}},
  90(430):443--450.

\bibitem[Bowden et~al., 2015]{bowden2015mendelian}
Bowden, J., Davey~Smith, G., and Burgess, S. (2015).
\newblock {Mendelian randomization with invalid instruments: effect estimation
  and bias detection through Egger regression}.
\newblock {\em {International Journal of Epidemiology}}, 44(2):512--525.

\bibitem[Bowden et~al., 2016a]{bowden2016consistent}
Bowden, J., Davey~Smith, G., Haycock, P.~C., and Burgess, S. (2016a).
\newblock {Consistent estimation in Mendelian randomization with some invalid
  instruments using a weighted median estimator}.
\newblock {\em {Genetic Epidemiology}}, 40(4):304--314.

\bibitem[Bowden et~al., 2017]{bowden2017framework}
Bowden, J., Del Greco~M, F., Minelli, C., Davey~Smith, G., Sheehan, N., and
  Thompson, J. (2017).
\newblock {A framework for the investigation of pleiotropy in two-sample
  summary data Mendelian randomization}.
\newblock {\em {Statistics in Medicine}}, 36(11):1783--1802.

\bibitem[Bowden et~al., 2016b]{bowden2016assessing}
Bowden, J., Del Greco~M, F., Minelli, C., Davey~Smith, G., Sheehan, N.~A., and
  Thompson, J.~R. (2016b).
\newblock {Assessing the suitability of summary data for two-sample Mendelian
  randomization analyses using MR-Egger regression: the role of the I 2
  statistic}.
\newblock {\em {International Journal of Epidemiology}}, 45(6):1961--1974.

\bibitem[Bowden et~al., 2019]{bowden2019improving}
Bowden, J., Del Greco~M, F., Minelli, C., Zhao, Q., Lawlor, D.~A., Sheehan,
  N.~A., Thompson, J., and Davey~Smith, G. (2019).
\newblock {Improving the accuracy of two-sample summary-data Mendelian
  randomization: moving beyond the NOME assumption}.
\newblock {\em {International Journal of Epidemiology}}, 48(3):728--742.

\bibitem[Bowden et~al., 2018]{bowden2018improving}
Bowden, J., Spiller, W., Del Greco~M, F., Sheehan, N., Thompson, J., Minelli,
  C., and Davey~Smith, G. (2018).
\newblock {Improving the visualization, interpretation and analysis of
  two-sample summary data Mendelian randomization via the Radial plot and
  Radial regression}.
\newblock {\em {International Journal of Epidemiology}}, 47(4):1264--1278.

\bibitem[Burgess et~al., 2013]{burgess2013mendelian}
Burgess, S., Butterworth, A., and Thompson, S.~G. (2013).
\newblock {Mendelian randomization analysis with multiple genetic variants
  using summarized data}.
\newblock {\em {Genetic Epidemiology}}, 37(7):658--665.

\bibitem[Burgess et~al., 2016]{burgess2016combining}
Burgess, S., Dudbridge, F., and Thompson, S.~G. (2016).
\newblock {Combining information on multiple instrumental variables in
  Mendelian randomization: comparison of allele score and summarized data
  methods}.
\newblock {\em {Statistics in Medicine}}, 35(11):1880--1906.

\bibitem[Burgess et~al., 2015]{burgess2015using}
Burgess, S., Scott, R.~A., Timpson, N.~J., Smith, G.~D., Thompson, S.~G.,
  Consortium, E.-I., et~al. (2015).
\newblock {Using published data in Mendelian randomization: a blueprint for
  efficient identification of causal risk factors}.
\newblock {\em {European Journal of Epidemiology}}, 30(7):543--552.

\bibitem[Burgess and Thompson, 2011]{burgess2011bias}
Burgess, S. and Thompson, S.~G. (2011).
\newblock {Bias in causal estimates from Mendelian randomization studies with
  weak instruments}.
\newblock {\em {Statistics in Medicine}}, 30(11):1312--1323.

\bibitem[Choi et~al., 2018]{choi2018weak}
Choi, J., Gu, J., and Shen, S. (2018).
\newblock {Weak-instrument robust inference for two-sample instrumental
  variables regression}.
\newblock {\em {Journal of Applied Econometrics}}, 33(1):109--125.

\bibitem[Davey~Smith and Ebrahim, 2003]{davey2003mendelian}
Davey~Smith, G. and Ebrahim, S. (2003).
\newblock {``Mendelian randomization'': can genetic epidemiology contribute to
  understanding environmental determinants of disease?}
\newblock {\em {International Journal of Epidemiology}}, 32(1):1--22.

\bibitem[Davidson and MacKinnon, 2014]{davidson2014confidence}
Davidson, R. and MacKinnon, J.~G. (2014).
\newblock {Confidence sets based on inverting Anderson--Rubin tests}.
\newblock {\em {The Econometrics Journal}}, 17(2):S39--S58.

\bibitem[Dufour, 1997]{dufour1997some}
Dufour, J.-M. (1997).
\newblock {Some impossibility theorems in econometrics with applications to
  structural and dynamic models}.
\newblock {\em {Econometrica: Journal of the Econometric Society}}, pages
  1365--1387.

\bibitem[Guo et~al., 2018]{guo2018confidence}
Guo, Z., Kang, H., Tony~Cai, T., and Small, D.~S. (2018).
\newblock {Confidence intervals for causal effects with invalid instruments by
  using two-stage hard thresholding with voting}.
\newblock {\em {Journal of the Royal Statistical Society: Series B (Statistical
  Methodology)}}, 80(4):793--815.

\bibitem[Hartwig et~al., 2017]{hartwig2017robust}
Hartwig, F.~P., Davey~Smith, G., and Bowden, J. (2017).
\newblock {Robust inference in summary data Mendelian randomization via the
  zero modal pleiotropy assumption}.
\newblock {\em {International Journal of Epidemiology}}, 46(6):1985--1998.

\bibitem[Hemani et~al., 2018]{hemani2018mr}
Hemani, G., Zheng, J., Elsworth, B., Wade, K.~H., Haberland, V., Baird, D.,
  Laurin, C., Burgess, S., Bowden, J., Langdon, R., et~al. (2018).
\newblock {The MR-Base platform supports systematic causal inference across the
  human phenome}.
\newblock {\em {Elife}}, 7:e34408.

\bibitem[Holland, 1988]{holland1988causal}
Holland, P.~W. (1988).
\newblock {Causal inference, path analysis and recursive structural equations
  models}.
\newblock {\em {ETS Research Report Series}}, 1988(1):i--50.

\bibitem[Johnson et~al., 2008]{johnson2008snap}
Johnson, A.~D., Handsaker, R.~E., Pulit, S.~L., Nizzari, M.~M., O'Donnell,
  C.~J., and De~Bakker, P.~I. (2008).
\newblock {SNAP: a web-based tool for identification and annotation of proxy
  SNPs using HapMap}.
\newblock {\em {Bioinformatics}}, 24(24):2938--2939.

\bibitem[Kang et~al., 2020]{kang2020two}
Kang, H., Lee, Y., Cai, T.~T., and Small, D.~S. (2020).
\newblock {Two robust tools for inference about causal effects with invalid
  instruments}.
\newblock {\em {Biometrics}}.

\bibitem[Kang et~al., 2016]{kang2016instrumental}
Kang, H., Zhang, A., Cai, T.~T., and Small, D.~S. (2016).
\newblock {Instrumental variables estimation with some invalid instruments and
  its application to Mendelian randomization}.
\newblock {\em {Journal of the American Statistical Association}},
  111(513):132--144.

\bibitem[Kleibergen, 2002]{kleibergen2002pivotal}
Kleibergen, F. (2002).
\newblock {Pivotal statistics for testing structural parameters in instrumental
  variables regression}.
\newblock {\em {Econometrica}}, 70(5):1781--1803.

\bibitem[Lawlor et~al., 2008]{lawlor2008mendelian}
Lawlor, D.~A., Harbord, R.~M., Sterne, J.~A., Timpson, N., and Davey~Smith, G.
  (2008).
\newblock {Mendelian randomization: using genes as instruments for making
  causal inferences in epidemiology}.
\newblock {\em {Statistics in Medicine}}, 27(8):1133--1163.

\bibitem[Locke et~al., 2015]{locke2015genetic}
Locke, A.~E., Kahali, B., Berndt, S.~I., Justice, A.~E., Pers, T.~H., Day,
  F.~R., Powell, C., Vedantam, S., Buchkovich, M.~L., Yang, J., et~al. (2015).
\newblock {Genetic studies of body mass index yield new insights for obesity
  biology}.
\newblock {\em {Nature}}, 518(7538):197.

\bibitem[Moreira and Moreira, 2019]{moreira2019optimal}
Moreira, H. and Moreira, M.~J. (2019).
\newblock {Optimal two-sided tests for instrumental variables regression with
  heteroskedastic and autocorrelated errors}.
\newblock {\em {Journal of Econometrics}}, 213(2):398--433.

\bibitem[Moreira, 2003]{moreira2003conditional}
Moreira, M.~J. (2003).
\newblock {A conditional likelihood ratio test for structural models}.
\newblock {\em {Econometrica}}, 71(4):1027--1048.

\bibitem[Pierce and Burgess, 2013]{pierce2013efficient}
Pierce, B.~L. and Burgess, S. (2013).
\newblock {Efficient design for Mendelian randomization studies: subsample and
  2-sample instrumental variable estimators}.
\newblock {\em {American Journal of Epidemiology}}, 178(7):1177--1184.

\bibitem[Small and Rosenbaum, 2008]{small2008war}
Small, D.~S. and Rosenbaum, P.~R. (2008).
\newblock {War and wages: the strength of instrumental variables and their
  sensitivity to unobserved biases}.
\newblock {\em {Journal of the American Statistical Association}},
  103(483):924--933.

\bibitem[Staiger and Stock, 1997]{staiger65stock}
Staiger, D. and Stock, J.~H. (1997).
\newblock {Instrumental variables regression with weak instruments}.
\newblock {\em {Econometrica}}, 65:557--586.

\bibitem[Stock et~al., 2002]{stock2002survey}
Stock, J.~H., Wright, J.~H., and Yogo, M. (2002).
\newblock {A survey of weak instruments and weak identification in generalized
  method of moments}.
\newblock {\em {Journal of Business \& Economic Statistics}}, 20(4):518--529.

\bibitem[Windmeijer, 2019]{windmeijer2019two}
Windmeijer, F. (2019).
\newblock {Two-stage least squares as minimum distance}.
\newblock {\em {The Econometrics Journal}}, 22(1):1--9.

\bibitem[Yavorska and Burgess, 2017]{yavorska2017mendelianrandomization}
Yavorska, O.~O. and Burgess, S. (2017).
\newblock {MendelianRandomization: an R package for performing Mendelian
  randomization analyses using summarized data}.
\newblock {\em {International Journal of Epidemiology}}, 46(6):1734--1739.

\bibitem[Ye et~al., 2019]{ye2019debiased}
Ye, T., Shao, J., and Kang, H. (2019).
\newblock {Debiased inverse-variance weighted estimator in two-sample
  summary-data mendelian randomization}.
\newblock {\em {arXiv preprint arXiv:1911.09802}}.

\bibitem[Zhao, 2018]{mrrapsSoftware}
Zhao, Q. (2018).
\newblock mr.raps: Two sample mendelian randomization using robust adjusted
  profile score.
\newblock {\em CRAN}, (Version 0.2).

\bibitem[Zhao et~al., 2019]{zhao2017two}
Zhao, Q., Wang, J., Bowden, J., and Small, D.~S. (2019).
\newblock {Two-sample instrumental variable analyses using heterogeneous
  samples}.
\newblock {\em {Statistical Science}}, 34(2):317--333.

\bibitem[Zhao et~al., 2020]{zhao2020statistical}
Zhao, Q., Wang, J., Hemani, G., Bowden, J., Small, D.~S., et~al. (2020).
\newblock {Statistical inference in two-sample summary-data Mendelian
  randomization using robust adjusted profile score}.
\newblock {\em {Annals of Statistics}}, 48(3):1742--1769.

\end{thebibliography}


\begin{thebibliography}{}

\bibitem[Bowden et~al., 2016]{bowden2016assessing}
Bowden, J., Del Greco~M, F., Minelli, C., Davey~Smith, G., Sheehan, N.~A., and
  Thompson, J.~R. (2016).
\newblock {Assessing the suitability of summary data for two-sample Mendelian
  randomization analyses using MR-Egger regression: the role of the I 2
  statistic}.
\newblock {\em {International Journal of Epidemiology}}, 45(6):1961--1974.

\bibitem[Bowden et~al., 2019]{bowden2019improving}
Bowden, J., Del Greco~M, F., Minelli, C., Zhao, Q., Lawlor, D.~A., Sheehan,
  N.~A., Thompson, J., and Davey~Smith, G. (2019).
\newblock {Improving the accuracy of two-sample summary-data Mendelian
  randomization: moving beyond the NOME assumption}.
\newblock {\em {International Journal of Epidemiology}}, 48(3):728--742.

\bibitem[Bowden et~al., 2018]{bowden2018improving}
Bowden, J., Spiller, W., Del Greco~M, F., Sheehan, N., Thompson, J., Minelli,
  C., and Davey~Smith, G. (2018).
\newblock {Improving the visualization, interpretation and analysis of
  two-sample summary data Mendelian randomization via the Radial plot and
  Radial regression}.
\newblock {\em {International Journal of Epidemiology}}, 47(4):1264--1278.

\bibitem[Koller and Stahel, 2011]{koller2011sharpening}
Koller, M. and Stahel, W.~A. (2011).
\newblock {Sharpening wald-type inference in robust regression for small
  samples}.
\newblock {\em {Computational Statistics \& Data Analysis}}, 55(8):2504--2515.

\bibitem[Lederer and K{\"u}chenhoff, 2006]{lederer2006short}
Lederer, W. and K{\"u}chenhoff, H. (2006).
\newblock {A short Introduction to the SIMEX and MCSIMEX}.
\newblock {\em {The Newsletter of the R Project Volume 6/4, October 2006}},
  6:1--26.

\bibitem[Yavorska and Burgess, 2017]{yavorska2017mendelianrandomization}
Yavorska, O.~O. and Burgess, S. (2017).
\newblock {MendelianRandomization: an R package for performing Mendelian
  randomization analyses using summarized data}.
\newblock {\em {International Journal of Epidemiology}}, 46(6):1734--1739.

\bibitem[Yohai, 1987]{yohai1987high}
Yohai, V.~J. (1987).
\newblock {High breakdown-point and high efficiency robust estimates for
  regression}.
\newblock {\em {The Annals of Statistics}}, pages 642--656.

\bibitem[Zhao et~al., 2020]{zhao2020statistical}
Zhao, Q., Wang, J., Hemani, G., Bowden, J., Small, D.~S., et~al. (2020).
\newblock {Statistical inference in two-sample summary-data Mendelian
  randomization using robust adjusted profile score}.
\newblock {\em {Annals of Statistics}}, 48(3):1742--1769.

\end{thebibliography}
\bibliographystyle{apalike}

\end{document}


\setstretch{1.54}

\title{Supporting Information for ``Weak-Instrument Robust Tests in Two-Sample Summary-Data Mendelian Randomization" by S. Wang and H. Kang}
\author{}
\date{}
\maketitle
\begin{abstract}
In Appendix A, we provide additional simulation results from the main manuscript. Appendix B contains proofs for the technical results in the main paper.
\end{abstract}

\setstretch{1.34}
\section{Web Appendix A: Extended Simulation Results}
{\color{black} \subsection{Extended Results: Size and Power of Proposed Tests When $r = 50$}



Figures \ref{size} and \ref{power} are the size and power results of the simulation study in Section 3.1 when $r= 50$. Similar to the results in the main paper, except for MR-RAPS, the size distortion of the pre-existing methods increases as the true exposure effect $\beta$ moves away from $0$. In contrast, our methods and MR-RAPS always maintain Type I error rate close to 5\%. For power, all the methods except MR-Egger.r have similar power under $H_0: \beta = 0$ (left panel). When $H_0: \beta = 1$, MR-RAPS, mrCLR, and mrK have identical power while maintaining a Type I error rate of 5\%. mrAR also maintains a Type I error rate of 5\%, but has slightly less power than MR-RAPS, mrCLR, and mrK. Finally, W.Median and MR-Egger.r have dramatic size distortions. Overall, we find that our proposed tests and MR-RAPS are superior to other pre-existing methods with respect to Type I error control and power when $r= 50$.

\begin{figure}
\begin{center}
\includegraphics[width=12cm]{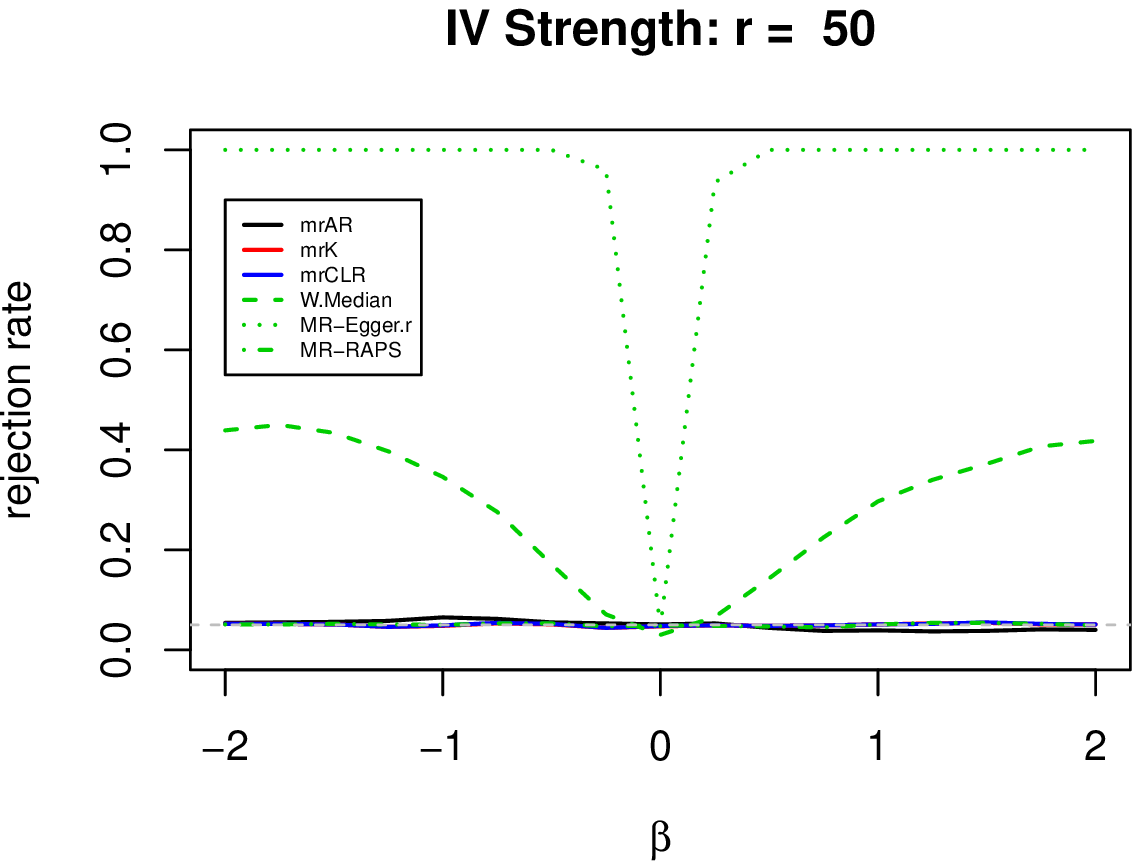}
\caption{Size under $r=50$. The number of instruments is $L=100$. $r$ represents instrument strength, with higher indicating stronger instruments; $r$ approximately corresponds to the first-stage F statistic for IV strength. The grey dashed horizontal line represents the Type I error rate of $0.05$.}
\label{size}
\end{center}
\end{figure}

\begin{figure}
\begin{center}
\includegraphics[width=16cm]{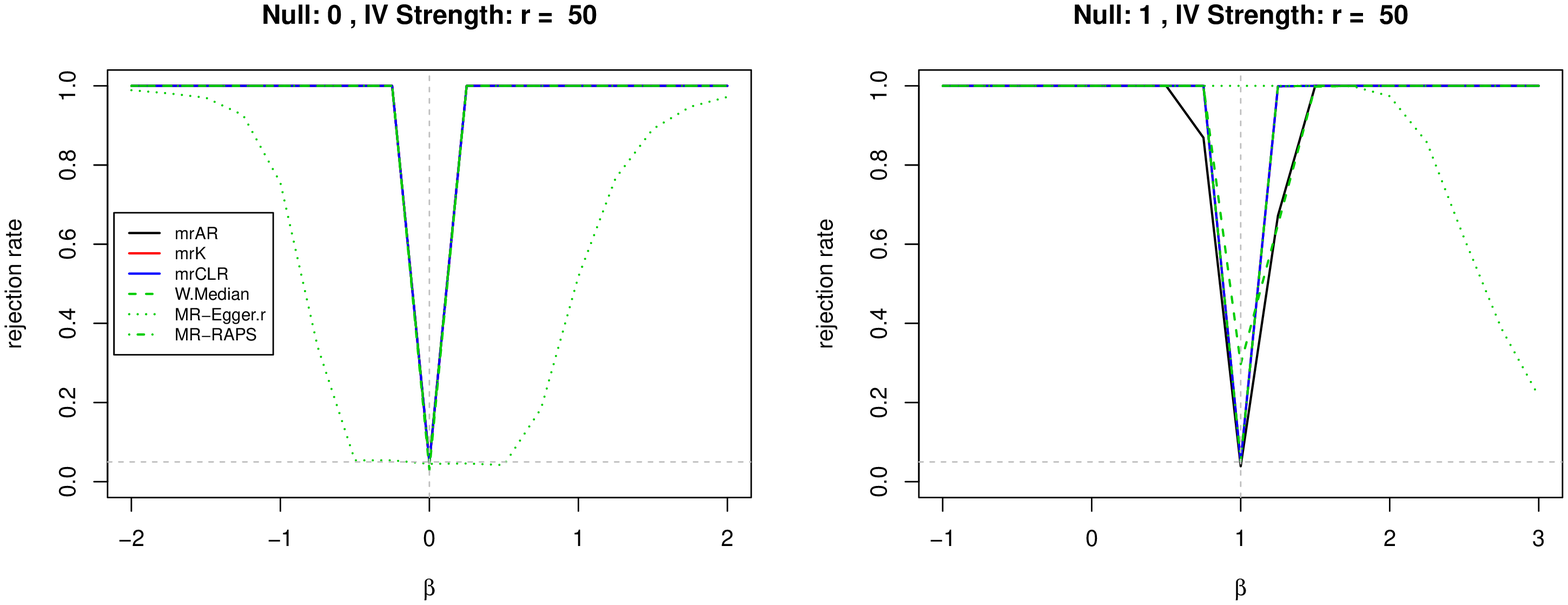}
\caption{Power curves under different nulls when $r = 50$. The left panel is under $H_0: \beta = 0$ and the right panel is under $H_0: \beta = 1$; each null value is represented by a grey dashed vertical line. $r$ represents instrument strength, with higher indicating stronger instruments; $r$ approximately corresponds to the first-stage F statistic for IV strength. The grey dashed horizontal line represents the Type I error rate of $0.05$.}
\label{power}
\end{center}
\end{figure}
}


{\color{black}
\subsection{Performance of Adjusted Tests under Correlated Instruments}
We extend the simulation study in Section 3.3 of the main manuscript and explore different mis-specifications of the correlation matrix. 

\subsubsection{Varying $r$}

We first examine the performance of our proposed tests when we vary instrument strength $r$. Specifically, we follow the same setup as in Section 3.3 of the main paper except we set $r = 4, 16$ and $25$.

Figures \ref{corr4},\ref{corr16}, and \ref{corr25} show the size and power of our adjusted tests. When the correlation $\hat{\rho}$ is under- or correctly-estimated ($\hat{\rho} \le 0.3$), the Type I error rate of our adjusted tests is slightly inflated. However, when $\hat{\rho}$ is over-estimated, the size distortion of mrAR is huge, while mrK and mrCLR produced inflated, but reasonable size control. All our adjusted tests always have power even with misspecified correlation $\hat{\rho}$ and their power improves with a larger $r$.

\begin{figure}
\centerline{\includegraphics[width = \textwidth]{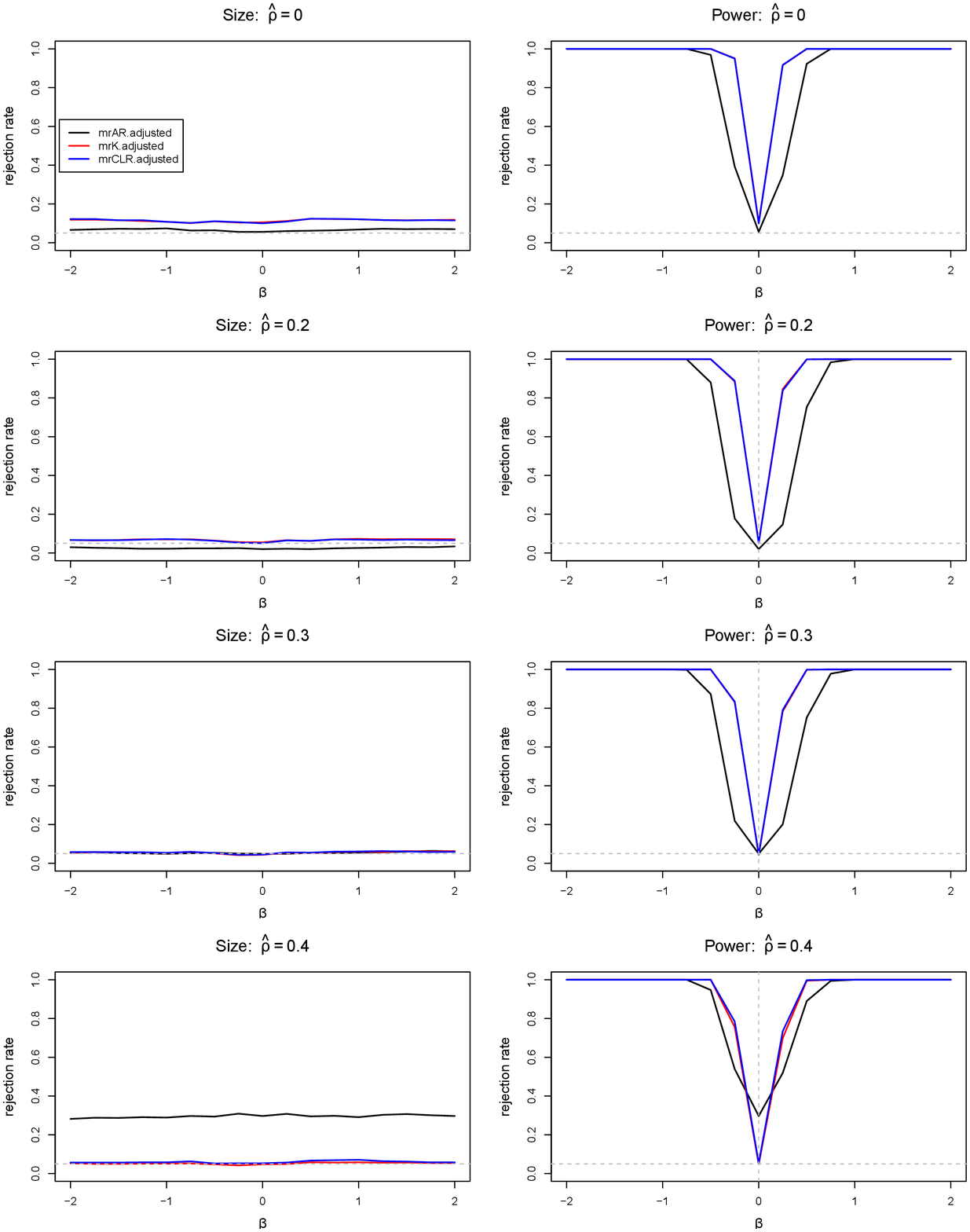}}
\caption{Performance of proposed tests with correlated instruments when $r=4$. The left panel shows size and the right panel shows power of testing $H_0: \beta = 0$. The true correlation $\rho$ is $0.3$ and each row changes the magnitude of $\hat{\rho}$ from $0$ to $0.4$. The true value of $\beta$ in the power plot is represented by a grey dashed vertical line at $0$. The grey dashed horizontal line represents the Type I error rate of $0.05$.}
\label{corr4}
\end{figure}

\begin{figure}
\centerline{\includegraphics[width = \textwidth]{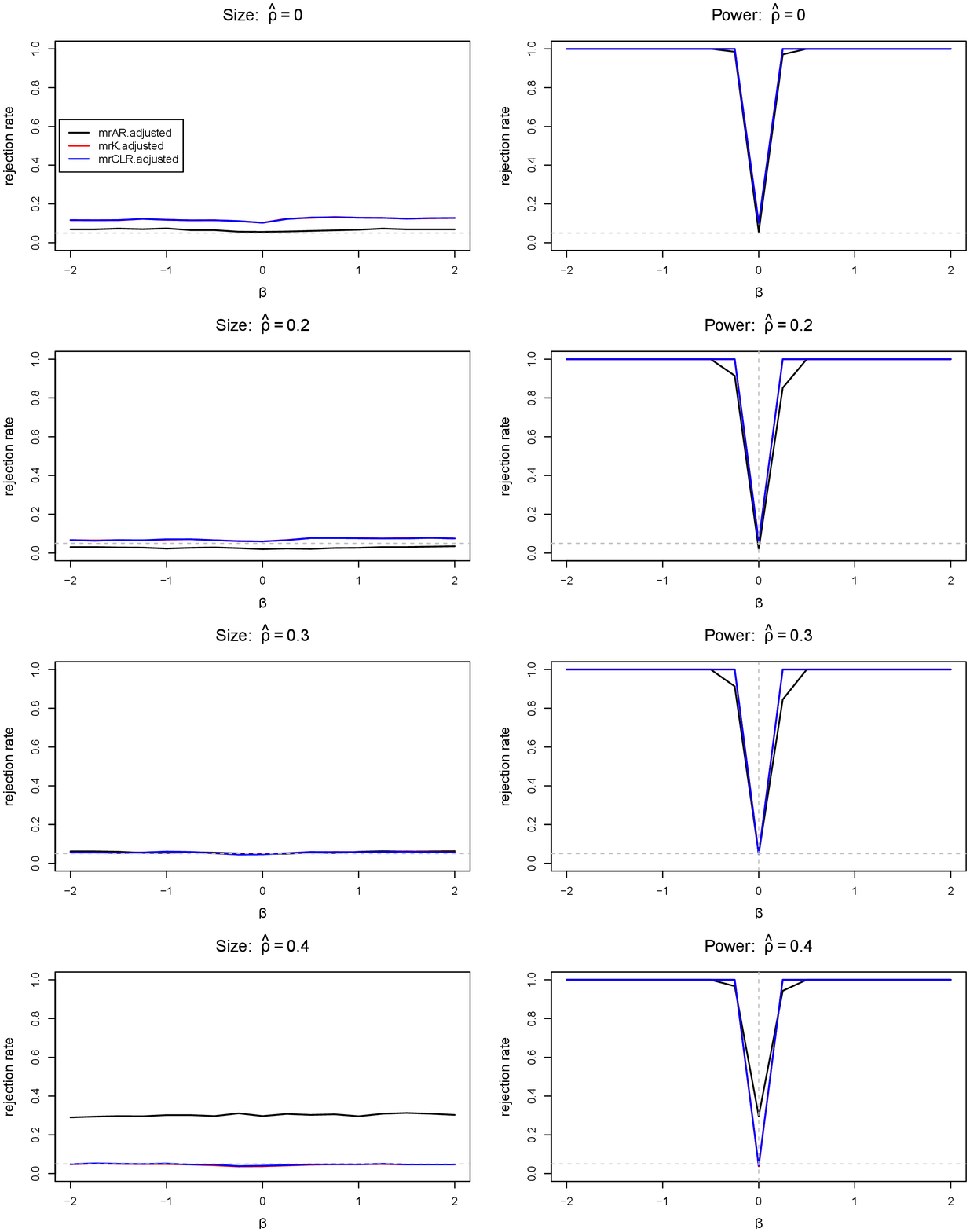}}
\caption{ Performance of proposed tests with correlated instruments when $r=16$. The left panel shows size and the right panel shows power of testing $H_0: \beta = 0$. The true correlation $\rho$ is $0.3$ and each row changes the magnitude of $\hat{\rho}$ from $0$ to $0.4$. The true value of $\beta$ in the power plot is represented by a grey dashed vertical line at $0$. The grey dashed horizontal line represents the Type I error rate of $0.05$.}
\label{corr16}
\end{figure}

\begin{figure}
\centerline{\includegraphics[width = \textwidth]{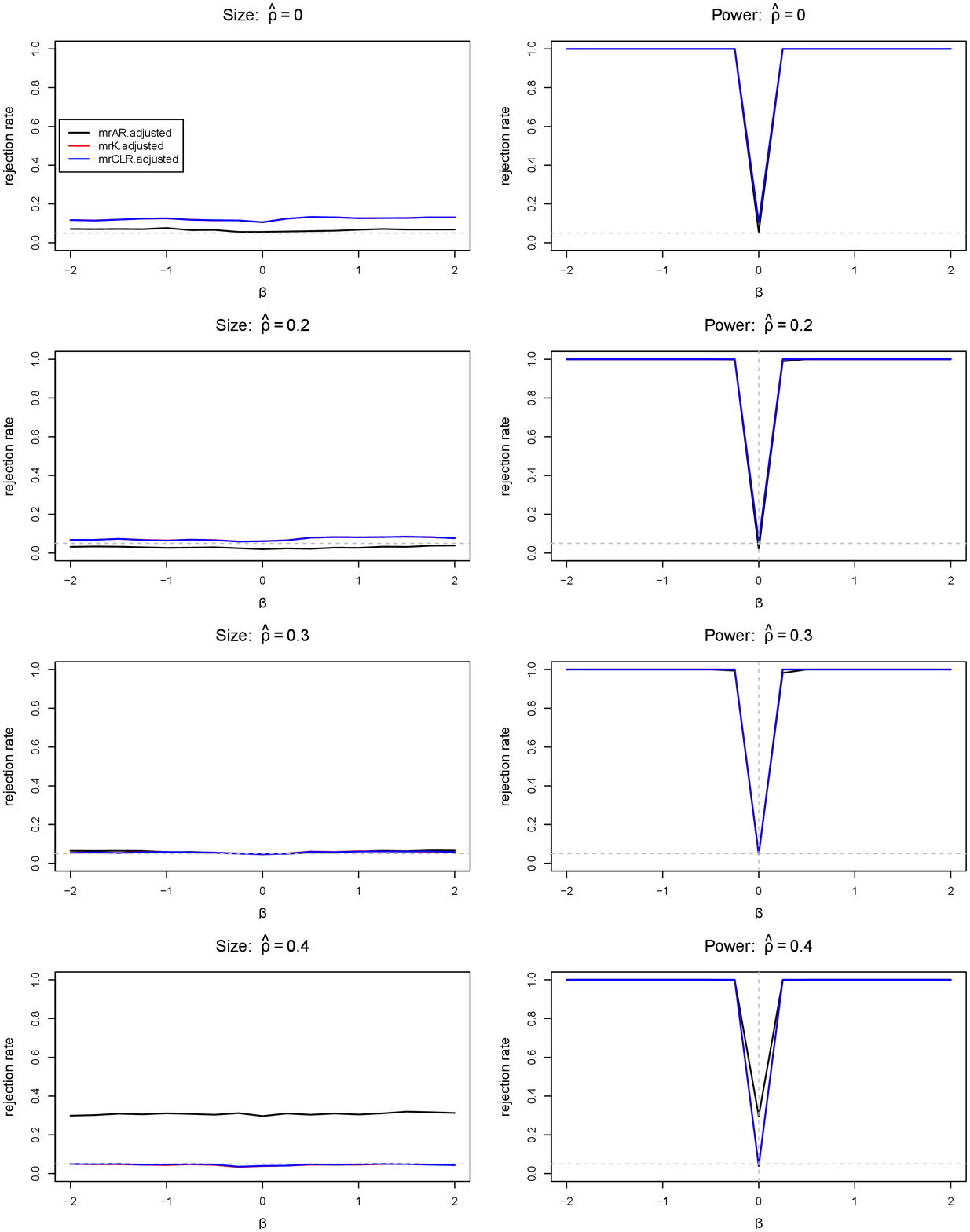}}
\caption{Performance of proposed tests with correlated instruments when $r=25$. The left panel shows size and the right panel shows power of testing $H_0: \beta = 0$. The true correlation $\rho$ is $0.3$ and each row changes the magnitude of $\hat{\rho}$ from $0$ to $0.4$. The true value of $\beta$ in the power plot is represented by a grey dashed vertical line at $0$. The grey dashed horizontal line represents the Type I error rate of $0.05$.}
\label{corr25}
\end{figure}

\subsubsection{Varying the Correlation Structure}


In this section, we vary how the instruments are correlated with each other by setting the bandwidth of the correlation matrix $M_{l}$ to be $w_l = 2$ and letting $\rho = 0.3$ if $|i - j| \le w_l$ and $\rho = 0$ otherwise. In other words, $w_l$ neighboring window of SNPs are correlated with each other. For the estimated $\widehat{M}_l$, we simulate cases when the working values are $\widehat{w_l} = 1, 2, 3$, and $\widehat{\rho} = 0, 0.2, 0.3, 0.4$. For all the simulations, we set $r=1$.

Figures \ref{corr2_1},\ref{corr2_2}, and \ref{corr2_3} show the power and size curves of our proposed tests. 
 From Figures \ref{corr2_1} and \ref{corr2_2}, we can see that our proposed tests have size control when $\rho$ is correctly or under-estimated. Also, our adjusted tests have power under those scenarios. However, when we over-estimate the bandwidth of the correlation matrix, the size of our tests, especially mrAR, exceeds $5\%$ even when we have an accurate estimate of the correlation $\rho$. In the worst scenario where we over-estimate both the bandwidth and the magnitude of correlation, the size distortion of our tests is huge. Overall, our proposed tests perform well when we have sufficient amount of information about the true correlation matrix, but suffer from serious size distortion when the bandwidth of the correlation matrix and the magnitude of the correlation are over-estimated (i.e. $\hat{w}_{l} >2$ and $\hat{\rho} > 0.3$).

\begin{figure}
\begin{center}
\includegraphics[width = 15cm]{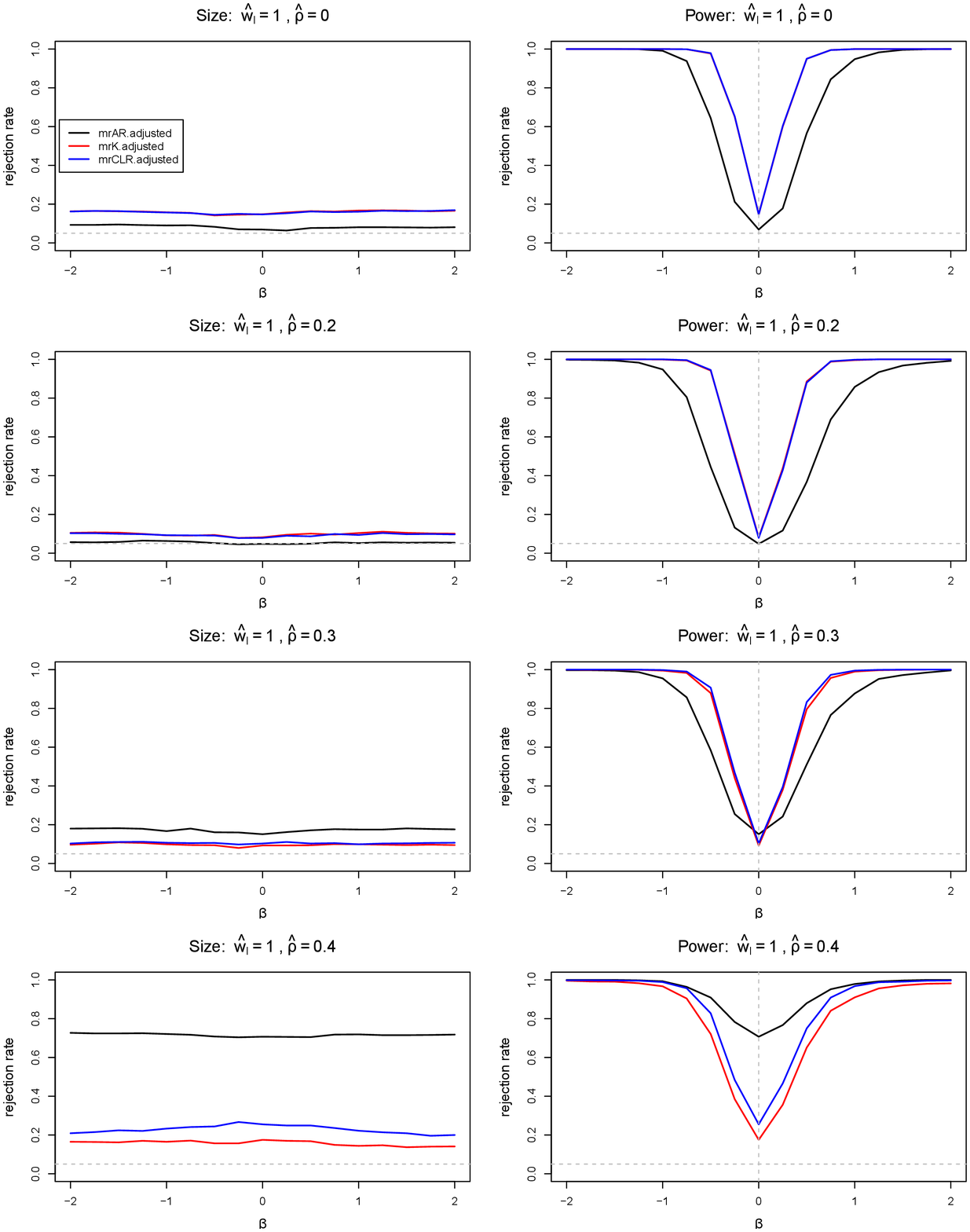}
\caption{Performance of proposed tests with correlated instruments when $r=1$. The left panel shows size and the right panel shows power of testing $H_0: \beta = 0$. The true correlation $\rho$ is $0.3$ and each row changes the magnitude of $\hat{\rho}$ from $0$ to $0.4$. The true bandwidth of the correlation matrix is $2$ and the working value of the bandwidth is $1$ throughout all plots. The true value of $\beta$ in the power plot is represented by a grey dashed vertical line at $0$. The grey dashed horizontal line represents the Type I error rate of $0.05$.}
\label{corr2_1}
\end{center}
\end{figure}

\begin{figure}
\includegraphics[width = 15cm]{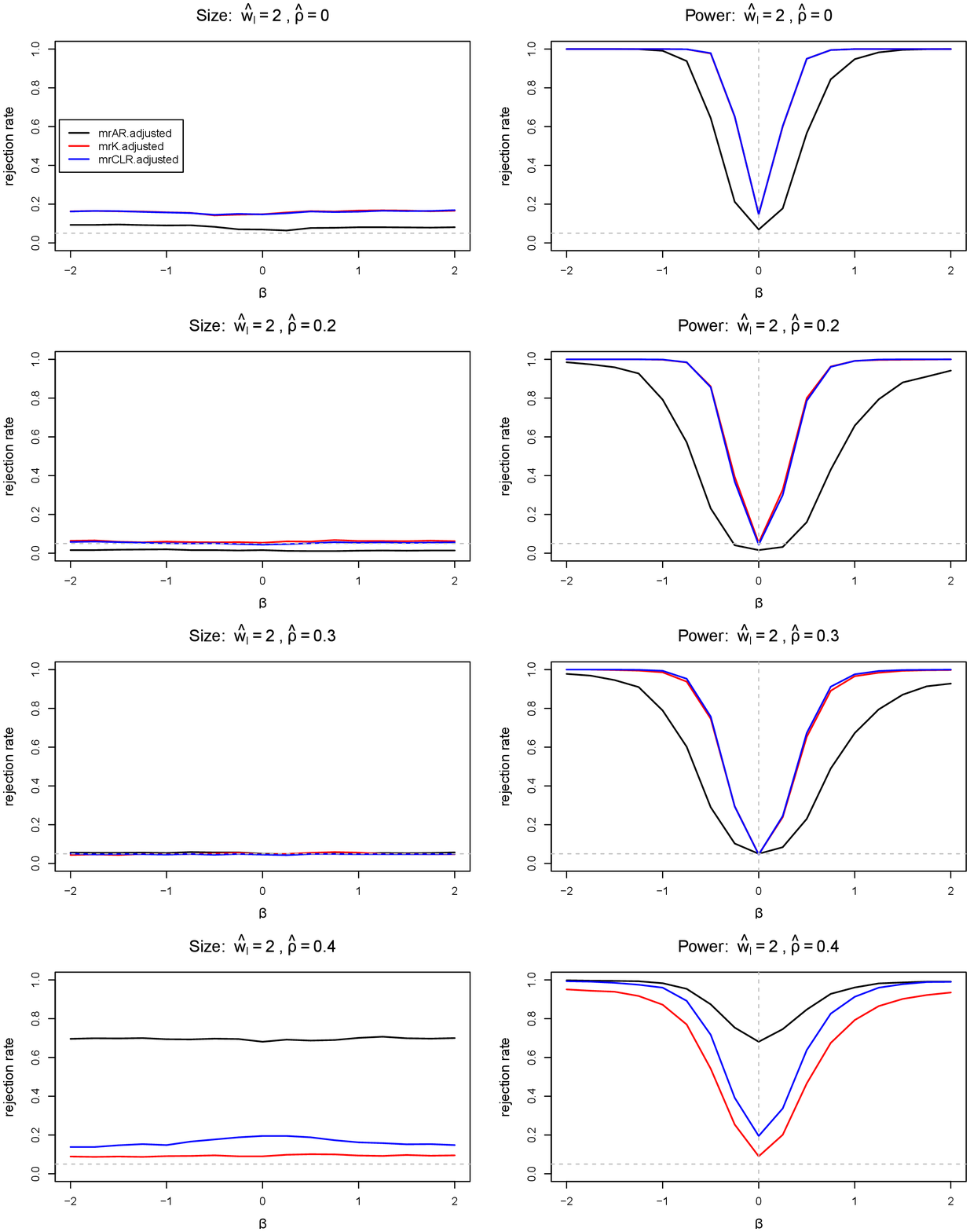}
\caption{ Performance of proposed tests with correlated instruments when $r=1$. The left panel shows size and the right panel shows power of testing $H_0: \beta = 0$. The true correlation $\rho$ is $0.3$ and each row changes the magnitude of $\hat{\rho}$ from $0$ to $0.4$. The true bandwidth of the correlation matrix is $2$ and the working value of the bandwidth is $2$ throughout all plots. The true value of $\beta$ in the power plot is represented by a grey dashed vertical line at $0$. The grey dashed horizontal line represents the Type I error rate of $0.05$.}
\label{corr2_2}
\end{figure}

\begin{figure}
\includegraphics[width = 15cm]{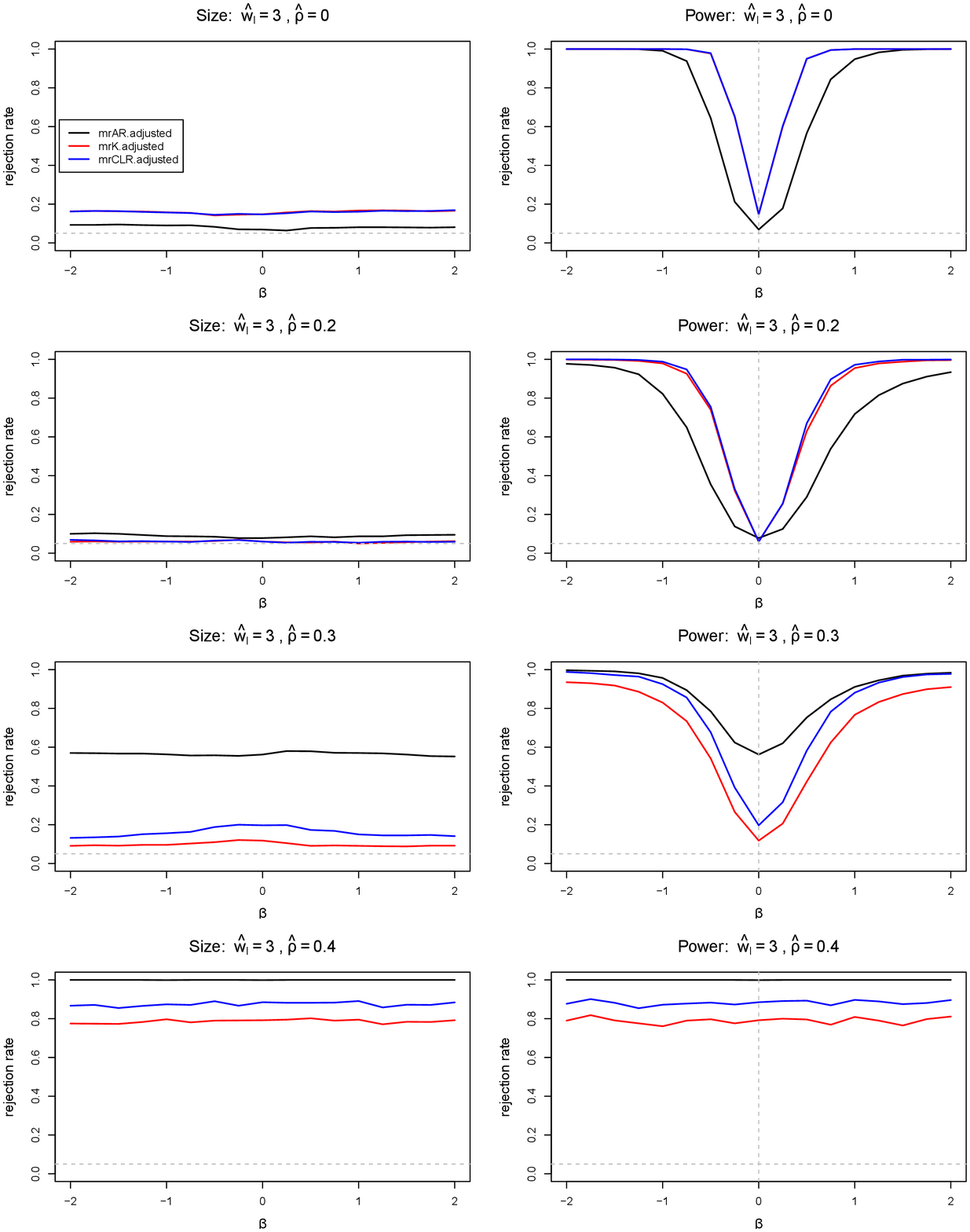}
\caption{Performance of proposed tests with correlated instruments when $r=1$. The left panel shows size and the right panel shows power of testing $H_0: \beta = 0$. The true correlation $\rho$ is $0.3$ and each row changes the magnitude of $\hat{\rho}$ from $0$ to $0.4$. The true bandwidth of the correlation matrix is $2$ and the working value of the bandwidth is $3$ throughout all plots. The true value of $\beta$ in the power plot is represented by a grey dashed vertical line at $0$. The grey dashed horizontal line represents the Type I error rate of $0.05$.}
\label{corr2_3}
\end{figure}
}

\begin{figure}
\centerline{\includegraphics[width = \textwidth]{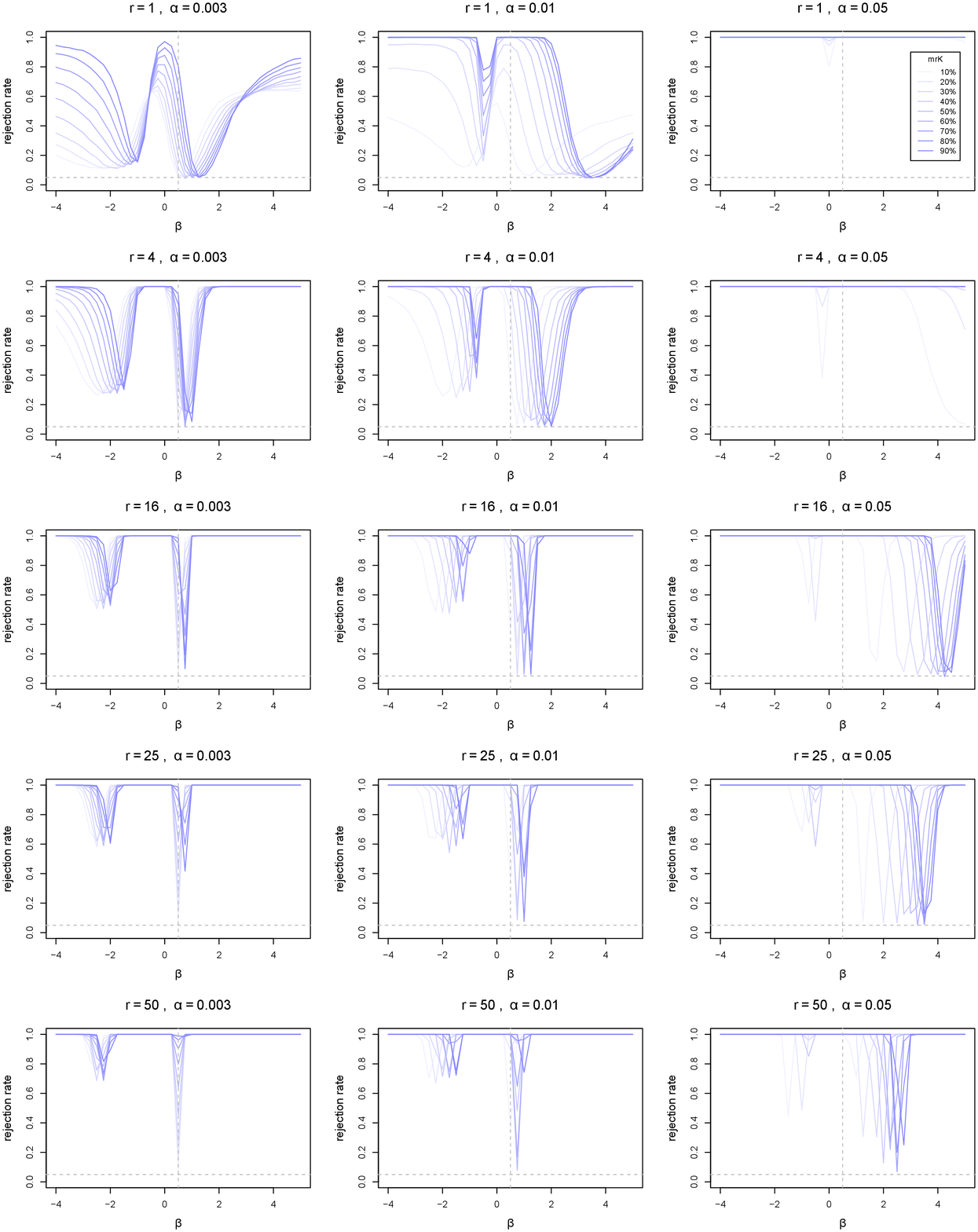}}
\caption{Rejection rates of mrK under invalid instruments. The true value of $\beta$ is represented by the vertical grey line at $0.5$ and we test $H_0: \beta = \beta_0$ for different values of $\beta_0$. $r$ represents instrument strength, with higher indicating stronger instruments; $r$ approximately corresponds to the first-stage F statistic for IV strength. Each column changes the magnitude of $\alpha$ from $0.003$ to $0.05$. Each purple or green line represents the proportion of invalid IVs as measured by the number of $\alpha_j \neq 0$ divided by $L$. The grey horizontal line represents the Type I error rate of $0.05$.}
\label{invalid_supp_mrK}
\end{figure}

\begin{figure}
\centerline{\includegraphics[width = \textwidth]{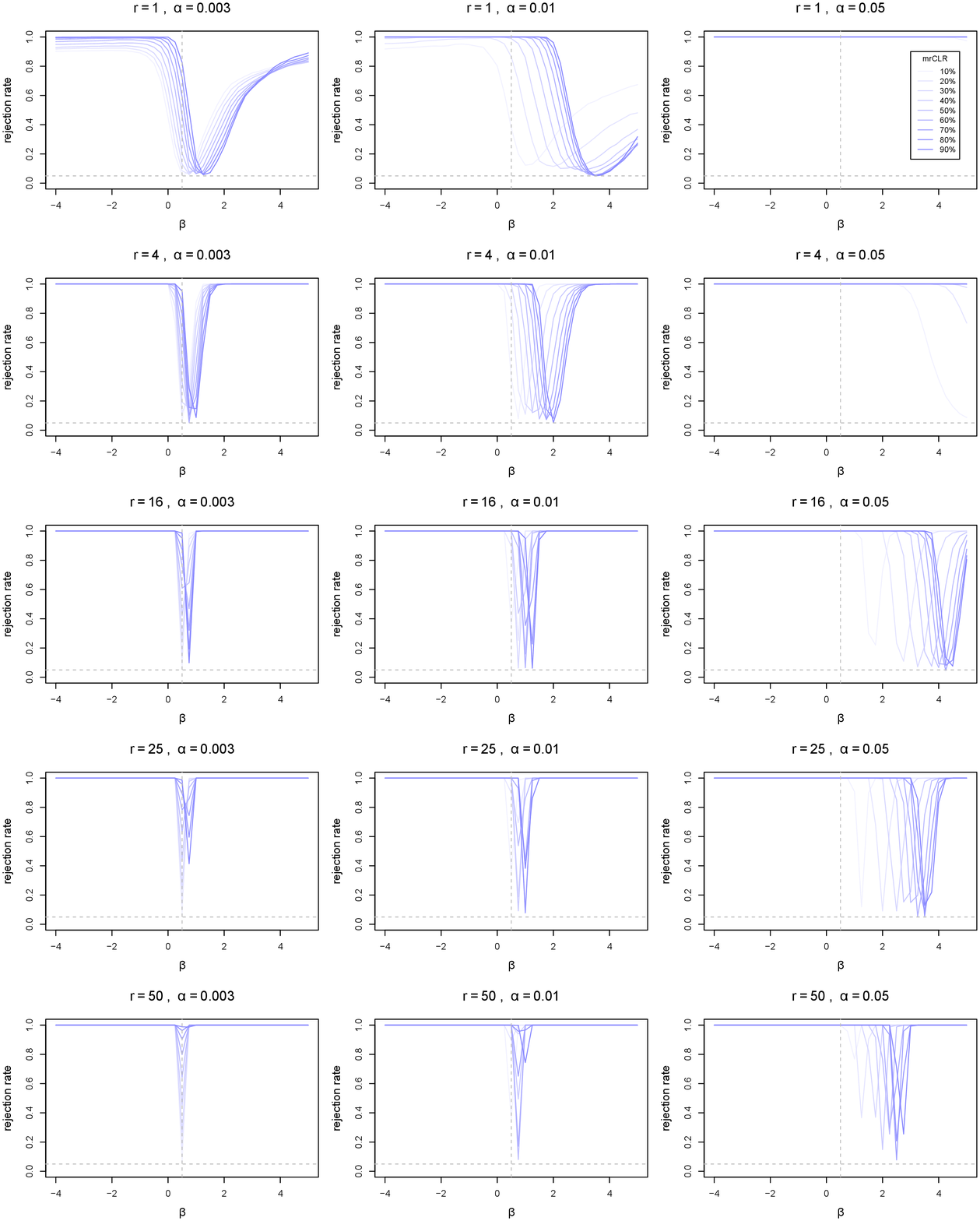}}
\caption{Rejection rates of mrCLR under invalid instruments. The true value of $\beta$ is represented by the vertical grey line at $0.5$ and we test $H_0: \beta = \beta_0$ for different values of $\beta_0$. $r$ represents instrument strength, with higher indicating stronger instruments; $r$ approximately corresponds to the first-stage F statistic for IV strength. Each column changes the magnitude of $\alpha$ from $0.003$ to $0.05$. Each purple or green line represents the proportion of invalid IVs as measured by the number of $\alpha_j \neq 0$ divided by $L$. The grey horizontal line represents the Type I error rate of $0.05$.}
\label{invalid_supp_mrCLR}
\end{figure}

\begin{figure}
\centerline{\includegraphics[width = \textwidth]{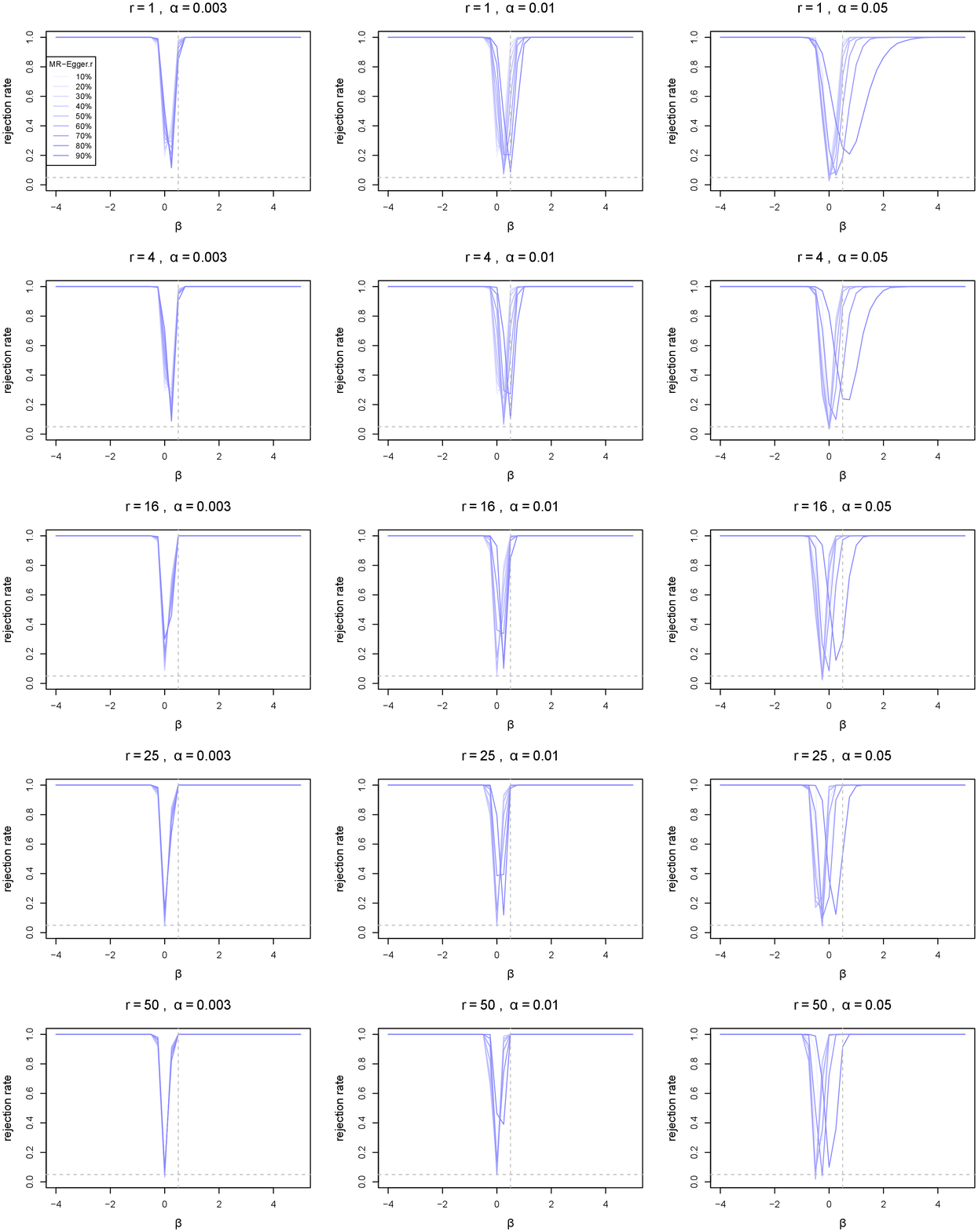}}
\caption{Rejection rates of MR-Egger.r under invalid instruments. The true value of $\beta$ is represented by the vertical grey line at $0.5$ and we test $H_0: \beta = \beta_0$ for different values of $\beta_0$. $r$ represents instrument strength, with higher indicating stronger instruments; $r$ approximately corresponds to the first-stage F statistic for IV strength. Each column changes the magnitude of $\alpha$ from $0.003$ to $0.05$. Each purple or green line represents the proportion of invalid IVs as measured by the number of $\alpha_j \neq 0$ divided by $L$. The grey horizontal line represents the Type I error rate set of $0.05$.}
\label{invalid_supp_mregger.r}
\end{figure}

\begin{figure}
\centerline{\includegraphics[width = \textwidth]{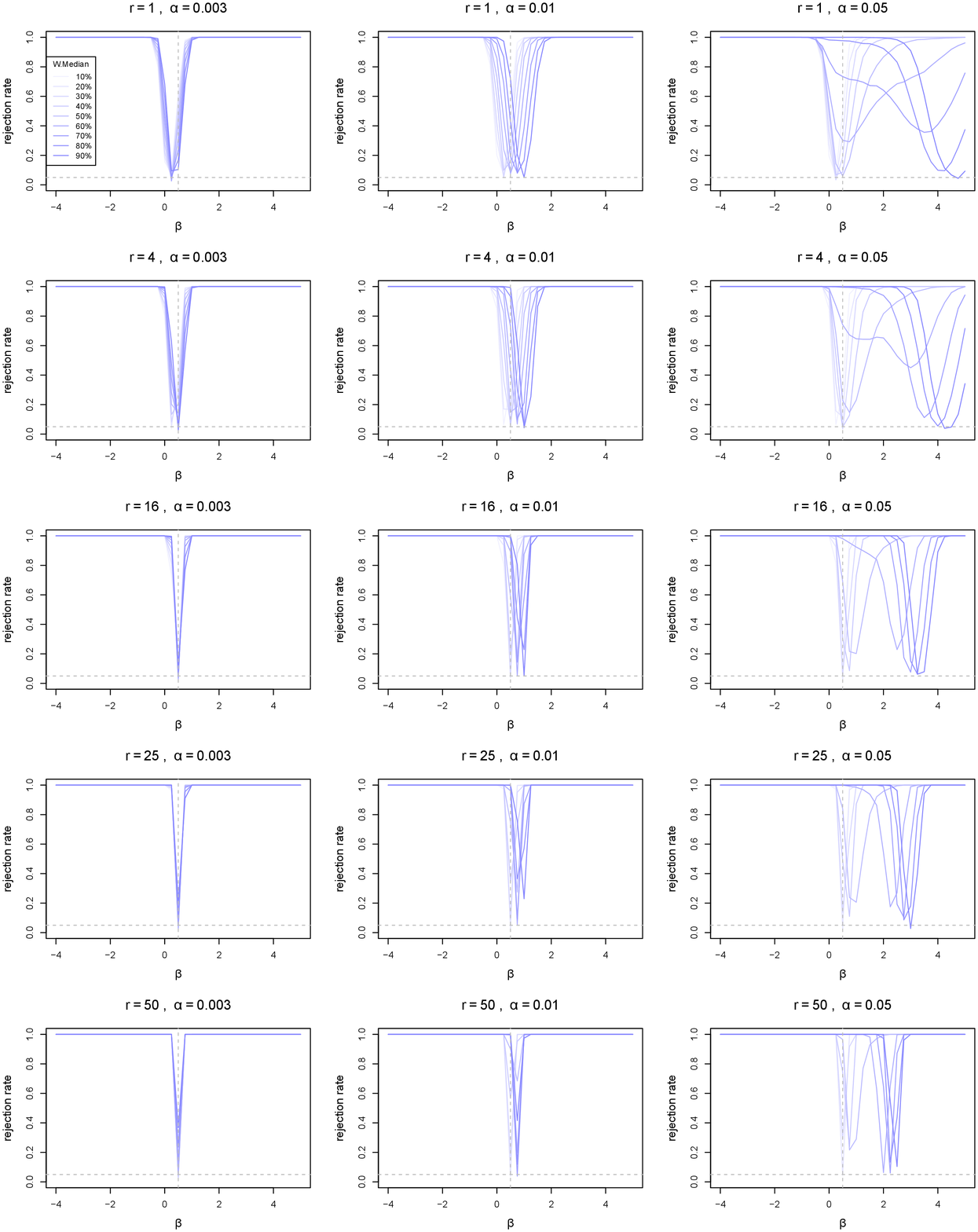}}
\caption{Rejection rates of W.Median under invalid instruments. The true value of $\beta$ is represented by the vertical grey line at $0.5$ and we test $H_0: \beta = \beta_0$ for different values of $\beta_0$. $r$ represents instrument strength, with higher indicating stronger instruments; $r$ approximately corresponds to the first-stage F statistic for IV strength. Each column changes the magnitude of $\alpha$ from $0.003$ to $0.05$. Each purple or green line represents the proportion of invalid IVs as measured by the number of $\alpha_j \neq 0$ divided by $L$. The grey horizontal line represents the Type I error rate set of $0.05$.}
\label{invalid_supp_wmedian}
\end{figure}

\begin{figure}
\centerline{\includegraphics[width = \textwidth]{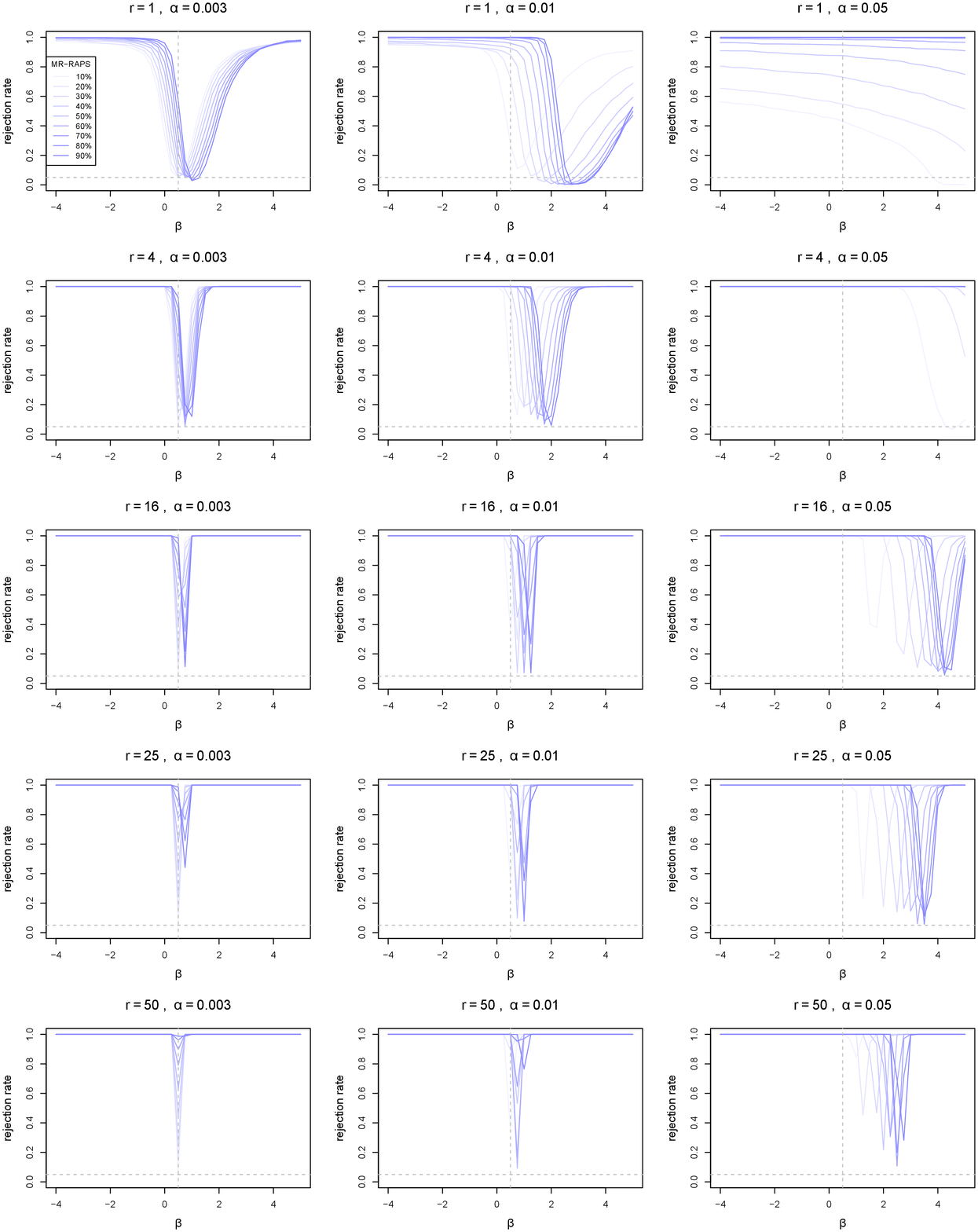}}
\caption{Rejection rates of MR-RAPS under invalid instruments. The true value of $\beta$ is represented by the vertical grey line at $0.5$ and we test $H_0: \beta = \beta_0$ for different values of $\beta_0$. $r$ represents instrument strength, with higher indicating stronger instruments; $r$ approximately corresponds to the first-stage F statistic for IV strength. Each column changes the magnitude of $\alpha$ from $0.003$ to $0.05$. Each purple or green line represents the proportion of invalid IVs as measured by the number of $\alpha_j \neq 0$ divided by $L$. The grey horizontal line represents the Type I error rate set of $0.05$.}
\label{invalid_supp_mrraps_simple}
\end{figure}

\begin{figure}
\centerline{\includegraphics[width = \textwidth]{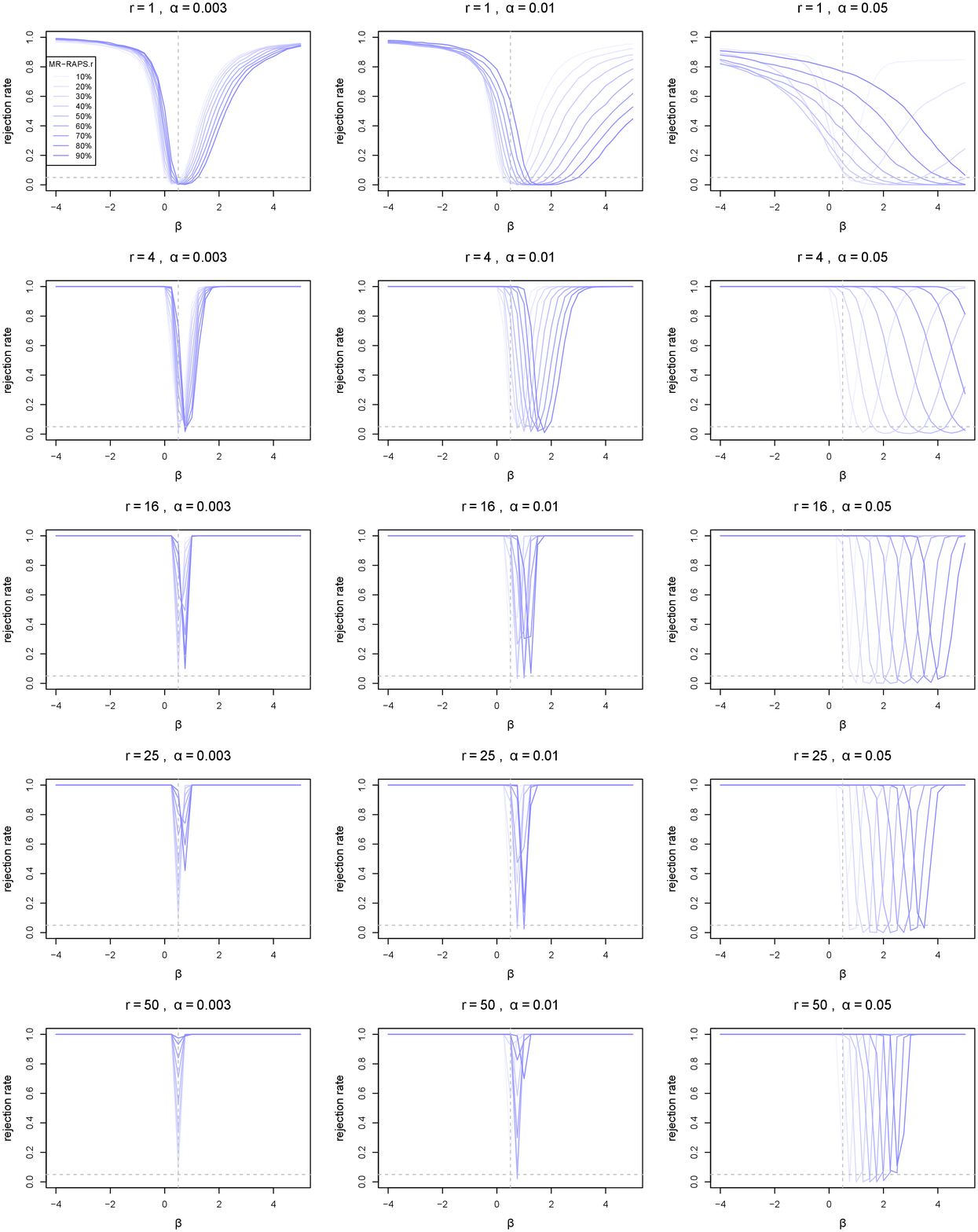}}
\caption{Rejection rates of MR-RAPS.r under invalid instruments. The true value of $\beta$ is represented by the vertical grey line at $0.5$ and we test $H_0: \beta = \beta_0$ for different values of $\beta_0$. $r$ represents instrument strength, with higher indicating stronger instruments; $r$ approximately corresponds to the first-stage F statistic for IV strength. Each column changes the magnitude of $\alpha$ from $0.003$ to $0.05$. Each purple or green line represents the proportion of invalid IVs as measured by the number of $\alpha_j \neq 0$ divided by $L$. The grey horizontal line represents the Type I error rate set of $0.05$.}
\label{invalid_supp_mrraps_robust}
\end{figure}

{\color{black}
\subsection{Performance of MR Methods Under Invalid Instruments}

Figure \ref{invalid_supp} shows the the simulation result in Section 3.2 of the main paper when $r= 50$. Similar to the results in the main paper, mrAR and the Q statistic remain above the grey horizontal line of $0.05$, alerting investigators about the presence of invalid instruments in the study. 

\begin{figure}
\centerline{\includegraphics[width = \textwidth]{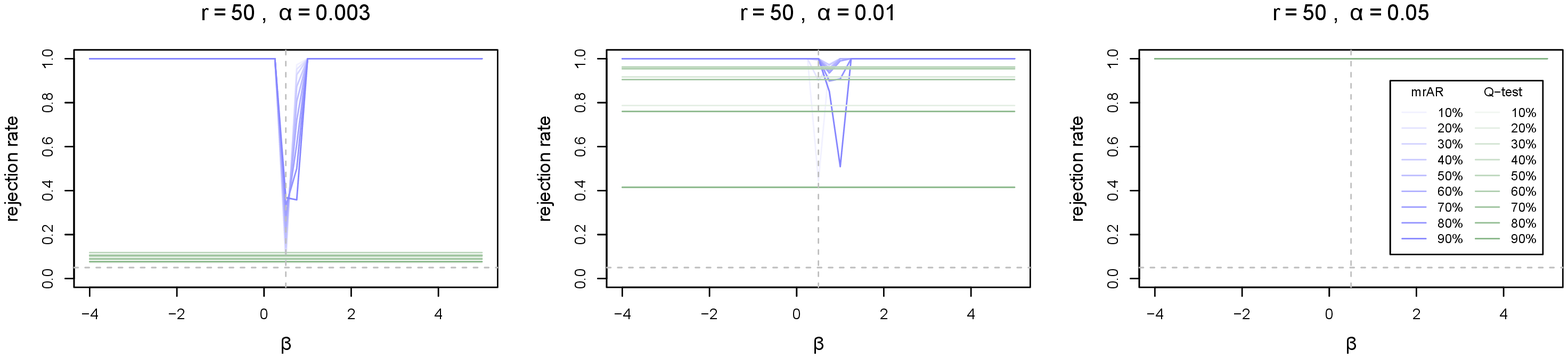}}
\caption{ Rejection rates of mrAR and Q-test (using modified second-order weights) under invalid instruments when $r= 50$. The true value of $\beta$ is represented by the vertical grey line at $0.5$ and we test $H_0: \beta = \beta_0$ for different values of $\beta_0$. $r$ represents instrument strength and approximately corresponds to the first-stage F statistic for IV strength. Each column changes the magnitude of $\alpha$ from $0.003$ to $0.05$. Each purple or green line represents the proportion of invalid IVs as measured by the number of $\alpha_j \neq 0$ divided by $L$. The grey horizontal line represents the Type I error rate of $0.05$.}
\label{invalid_supp}
\end{figure}

Figure \ref{invalid_Qtest2} shows the rejection rates of the Q statistic with exact modified weights. Overall, the Q statistic with modified second-order weights,  shows better power than the Q statistic using exact modified weights in this simulation study.

\begin{figure}
\centerline{\includegraphics[width = \textwidth]{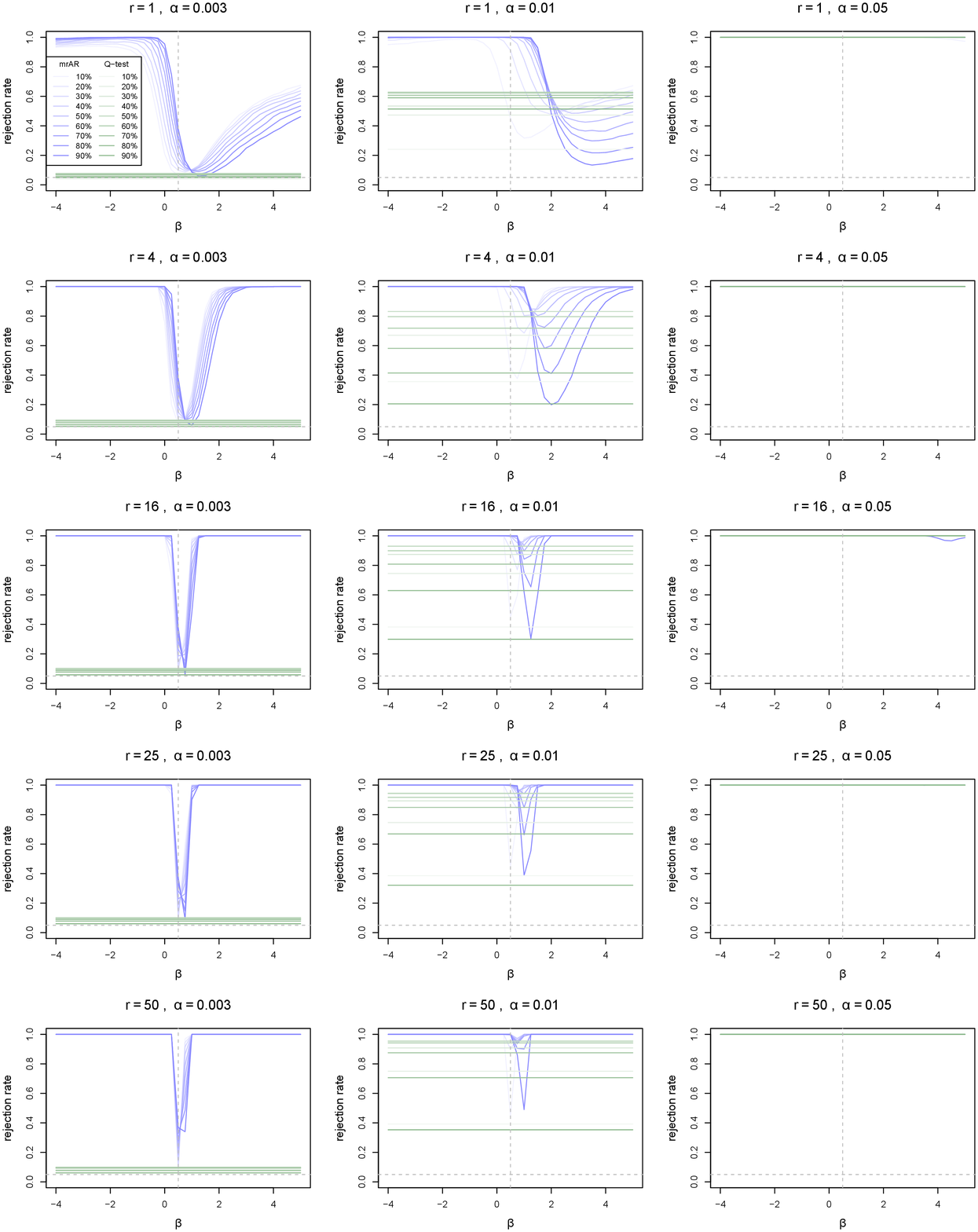}}
\caption{Rejection rates of mrAR and Q-test (using exact modified weights) under invalid instruments when $r= 50$. The true value of $\beta$ is represented by the vertical grey line at $0.5$ and we test $H_0: \beta = \beta_0$ for different values of $\beta_0$. $r$ represents instrument strength and approximately corresponds to the first-stage F statistic for IV strength. Each column changes the magnitude of $\alpha$ from $0.003$ to $0.05$. Each purple or green line represents the proportion of invalid IVs as measured by the number of $\alpha_j \neq 0$ divided by $L$. The grey horizontal line represents the Type I error rate of $0.05$.}
\label{invalid_Qtest2}
\end{figure}

Figures \ref{invalid_supp_mrK}, \ref{invalid_supp_mrCLR}, \ref{invalid_supp_mregger.r}, \ref{invalid_supp_wmedian}, \ref{invalid_supp_mrraps_simple}, and \ref{invalid_supp_mrraps_robust} repeat the simulation study in Section 3.2 of the main paper with mrK, mrCLR, MR-Egger.r, W.Median, and MR-RAPS.r, respectively. Not surprisingly, mrK and mrCLR are biased, with rejection rates dipping at or near $0.05$ not around the true value of $\beta = 0.5$. The bias increases with the magnitude of the direct effect and the proportion of invalid instruments, but decreases with $r$. Also, MR-Egger.r exhibits some bias because the InSIDE assumption is not satisfied. W.Median shows less bias when the proportion of invalid instruments is less than $50\%$ or when the magnitude of the violation is small. But, it becomes biased when more than $50\%$ of the instruments are violated or when $r$ is low, say around $1$ and $4$. MR-RAPS is unbiased when the magnitude of the direct effect is small ($\alpha = 0.003$) and the instruments are strong ($r = 50$), but it's biased away from the true causal effect as the direct effect increases. MR-RAPS.r performs well when the magnitude of the direct effect is small ($\alpha = 0.003$), but it generates biased confidence intervals when the magnitude of the direct effect increases ($\alpha = 0.01, 0.05$) across all values of $r$.

Overall, the simulation results from these existing methods echo many other existing works in the MR literature to compare different methods when faced with invalid instruments. Notably, no method, including ours, always show robust performance irrespective of the value of $r$ or the nature of invalid instruments and this may be an area for future work.}

\begin{figure}
\begin{center}
\includegraphics[width=16cm]{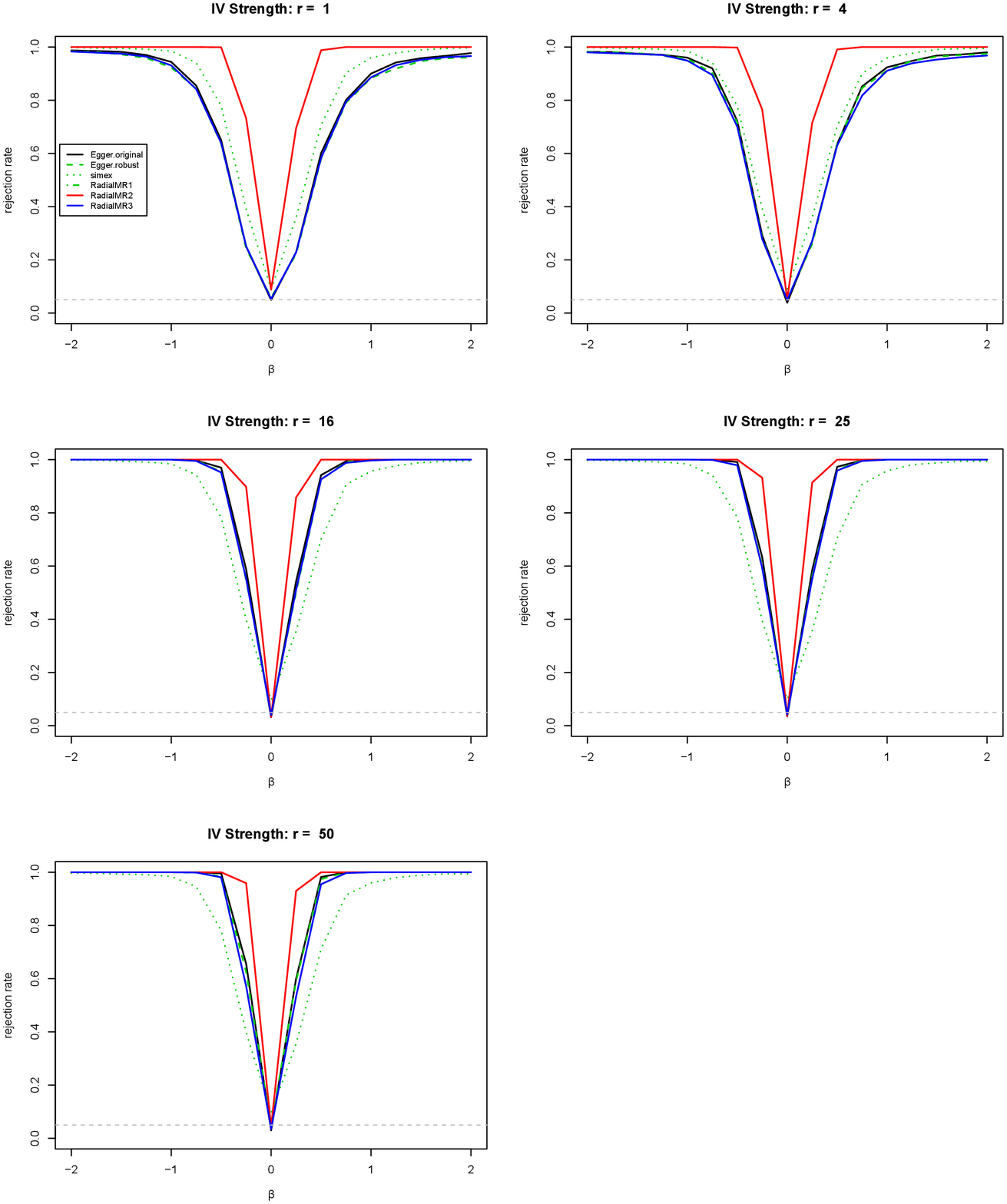}
\caption{Size of different MR-Egger methods under different IV strength. The number of instruments is $L=100$. $r$ represents instrument strength, with higher indicating stronger instruments; $r$ approximately corresponds to the first-stage F statistic for IV strength. The grey dashed horizontal line represents the Type I error rate set of $0.05$.}
\label{size_egger}
\end{center}
\end{figure}

\begin{figure}
\begin{center}
\includegraphics[width=14cm]{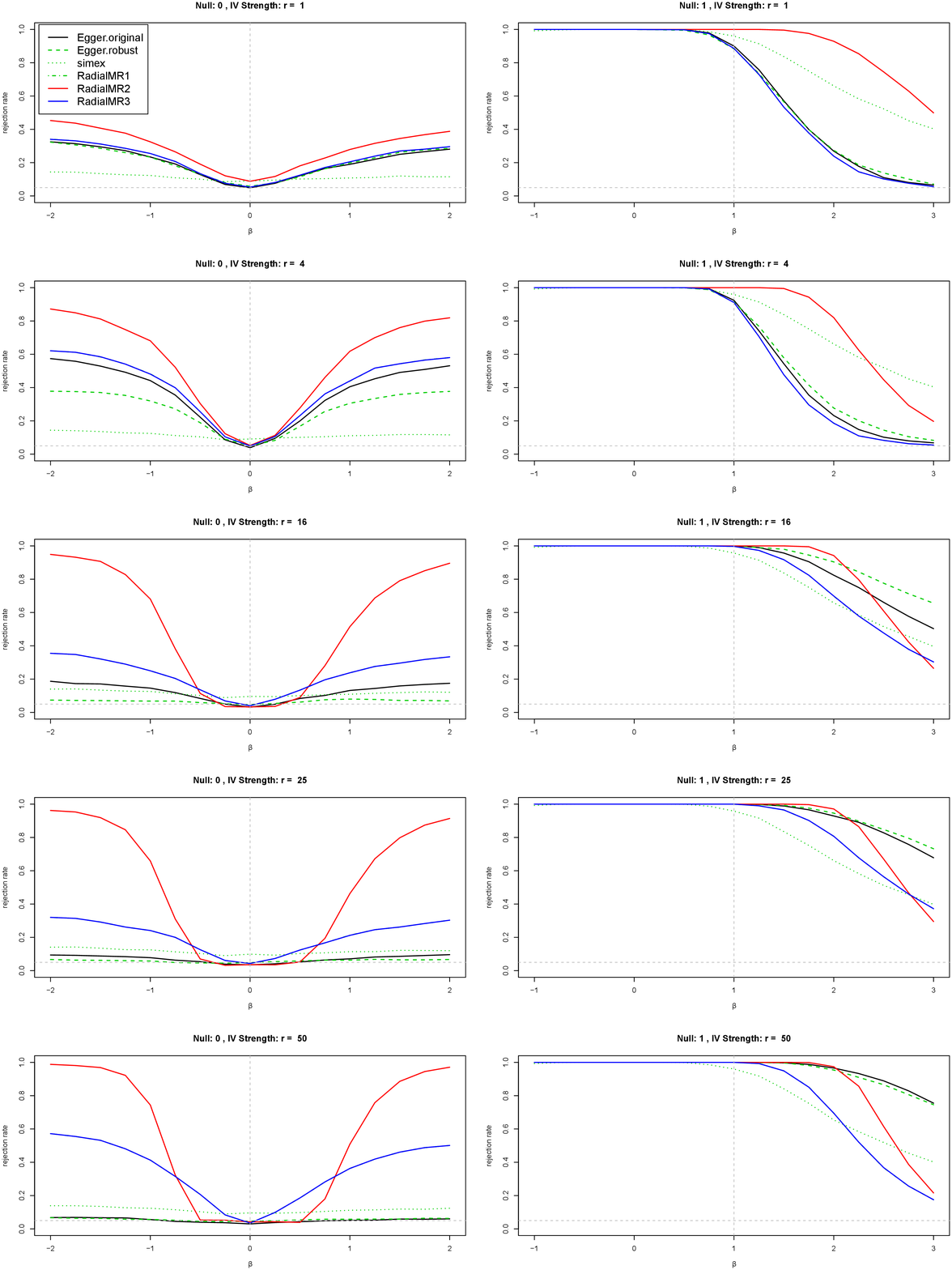}
\caption{Power curves of different MR-Egger methods under different nulls and IV strength. The left panel is under $H_0: \beta = 0$ and the right panel is under $H_0: \beta = 1$; each null value is represented by a grey dashed vertical line. $r$ represents instrument strength, with higher indicating stronger instruments; $r$ approximately corresponds to the first-stage F statistic for IV strength. The grey dashed horizontal line represents the Type I error rate set of $0.05$.}
\label{power_egger}
\end{center}
\end{figure}

{\color{black}
\subsection{Performance of Variants of MR-Egger Methods}
Here, we provide extended simulation results of six different variants of MR-Egger: the original MR-Egger regression (Egger.original), the robust MR-Egger regression via a robust regression (Egger.robust) \citep{yohai1987high, koller2011sharpening}, the robust MR-Egger via simulation extrapolation (simex) \citep{bowden2016assessing}, the robust MR-Egger via a radial regression using first-order weights (RadialMR1), second-order weights (RadialMR2), and modified second-order weights (RadialMR3) \citep{bowden2018improving}. We use the R package \verb+Mendelianrandomization+ \citep{yavorska2017mendelianrandomization}  to run Egger.original and Egger.robust, the R package \verb+simex+ \citep{lederer2006short} to run simex, and the R package \verb+RadialMR+ \citep{bowden2019improving} to run RadialMR1, RadialMR2, and RadialMR3.

We repeat the simulations in Section 3.1 of the main paper using all six variations of MR-Egger regression. Figure~\ref{size_egger} shows the size curves of all six MR-Egger methods. All the MR variations fail to control Type I error, but simex seems to have the smallest size distortions. Figure~\ref{power_egger} shows the power curves under the null hypothesis $H_0: \beta = 0$ and $H_0: \beta = 1$. Under the hypothesis of null effect $H_0: \beta = 0$, all the methods have Type I error control. However, under the null hypothesis $H_0: \beta = 1$, all the methods have rejection rate above $5\%$. In terms of power, RadialMR2 and RadialMR3 generally have better power than other MR-Egger variations. Also, simex has the lowest power across all settings. Based on the trade-offs between size and power, we opted to use RadialMR2 in the main text.}


\section{Web Appendix B: Proofs}
\subsection{Proof of Lemma 1.} 

Define
\begin{align*}
\tilde{S}(\beta_0) &= \left(\Sigma_{\Gamma}+ \beta_0^2 \Sigma_{\gamma} \right)^{-1/2} \left(\widehat{\Gamma}  - \beta_0 \widehat{\gamma} \right), \\
\tilde{R}(\beta_0) &= (\beta_0^2 \Sigma_{\Gamma}^{-1} + \Sigma_{\gamma}^{-1})^{-1/2} \left(\Sigma_{\Gamma}^{-1} \widehat{\Gamma}  \beta_0 +  \Sigma_{\gamma}^{-1} \hat{\gamma} \right) 
\end{align*}
Let $d_1 = \hat{\Gamma} - \Gamma, d_2 = \hat{\gamma} - \gamma$, then we have
\begin{align*}
\tilde{S}(\beta_0) &= (\Sigma_{\Gamma} + \beta_0^2 \Sigma_{\gamma})^{-1/2} [\hat{\Gamma}, \hat{\gamma}] b_0 \\
&= (n_1 \Sigma_{\Gamma} + n_1\beta_0^2 \Sigma_{\gamma})^{-1/2} n_1^{1/2} (I_L, -\beta_0 I_L)[vec(\Gamma, \gamma) + vec(d_1, d_2)]\\ 
&\inP (\Sigma_1 + c \beta_0^2 \Sigma_2)^{-1/2} (\beta - \beta_0)C + (\Sigma_{\Gamma} + \beta_0^2 \Sigma_{\gamma})^{-1/2}(I_L, -\beta_0 I_L)vec(d_1, d_2) \equiv S_\infty \\
&\sim N((\Sigma_1 + c \beta_0^2 \Sigma_2)^{-1/2} (\beta - \beta_0)C, I_L)
\end{align*}
\begin{align*}
\tilde{R}(\beta_0) &= (\beta_0^2 \Sigma_{\Gamma}^{-1} +  \Sigma_{\gamma}^{-1})^{-1/2} (\Sigma_{\Gamma}^{-1} \beta_0,  \Sigma_{\gamma}^{-1})vec(\hat{\Gamma}, \hat{\gamma})\\
&= (\beta_0^2 n_1^{-1} \Sigma_{\Gamma}^{-1} +  n_1^{-1}  \Sigma_{\gamma}^{-1})^{-1/2} [n_1^{-1}\Sigma_{\Gamma}^{-1} \beta_0, n_1^{-1} \Sigma_{\gamma}^{-1}] n_1^{1/2}vec(\Gamma, \gamma) + (\beta_0^2 \Sigma_{\Gamma}^{-1} +  \Sigma_{\gamma}^{-1})^{-1/2}[\Sigma_{\Gamma}^{-1} \beta_0,  \Sigma_{\gamma}^{-1}]\\
&\quad vec(d_1, d_2)\\
\inP &(\beta_0^2 \Sigma_1^{-1} + c^{-1} \Sigma_2^{-1})^{-1/2} [ \beta_0 \beta \Sigma_1^{-1} C + c^{-1} \Sigma_2^{-1}C]  + (\beta_0^2 \Sigma_{n,\Gamma}^{-1} +  \Sigma_{n,\gamma}^{-1})^{-1/2} (\Sigma_{n,\Gamma}^{-1} \beta_0,  \Sigma_{n,\gamma}^{-1}) vec(d_1, d_2) \equiv R_\infty \\
&\sim N((\beta_0^2 \Sigma_1^{-1} + c^{-1}\Sigma_2^{-1})^{-1/2} ( \beta_0 \beta \Sigma_1^{-1} C + c^{-1} \Sigma_2^{-1} C), I_L)
\end{align*}
Since $n_1(\hat{\Sigma}_{\Gamma} - \Sigma_{\Gamma} ) \stackrel{p}{\longrightarrow} 0, n_2 (\hat{\Sigma}_{\gamma} - \Sigma_{\gamma}) \stackrel{p}{\longrightarrow} 0$, we have that $( (S(\beta_0), R(\beta_0)) - (S_\infty, R_\infty)) \stackrel{p}{\longrightarrow} 0.$
The asymptotic normal distributions of $S_n$ and $R_n$ are independent because they are non-stochastic functions of $(\Sigma_{\Gamma} + \beta_0^2 \Sigma_{\gamma})^{-1/2}(I_L, -\beta_0 I_L)vec(d_1, d_2)$ and $(\beta_0^2 \Sigma_{\Gamma}^{-1} + \Sigma_{\gamma}^{-1})^{-1/2} (\Sigma_{\Gamma}^{-1} \beta_0,  \Sigma_{\gamma}^{-1}) vec(d_1, d_2)$, and 
\begin{align*}
&\quad Cov((\Sigma_{\Gamma} + \beta_0^2 \Sigma_{\gamma})^{-1/2}(I_L',-\beta_0 I_L')vec(d_1, d_2), (\beta_0^2 \Sigma_{\Gamma}^{-1} +  \Sigma_{\gamma}^{-1})^{-1/2}(\Sigma_{\Gamma}^{-1} \beta_0,  \Sigma_{\gamma}^{-1})vec(d_1, d_2)) \\
&= (\Sigma_{\Gamma} + \beta_0^2 \Sigma_{\gamma})^{-1/2}(I_L,-\beta_0 I_L) \left(
\begin{array}{ccccccc}
 \Sigma_{\Gamma} & 0\\
 0             & \Sigma_{\gamma}
\end{array}
\right)(\Sigma_{\Gamma}^{-1} \beta_0,  \Sigma_{\gamma}^{-1})'(\beta_0^2 \Sigma_{\Gamma}^{-1} +  \Sigma_{\gamma}^{-1})^{-1/2} \\
&= 0.
\end{align*}

\subsection{Proof of Theorem 1.} 

Let $Q_\infty = \left(
\begin{array}{ccccccc}
 Q_{\infty,S} & Q_{\infty,RS}\\
 Q_{\infty,SR} & Q_{\infty,R}
\end{array}
\right) = \left(
\begin{array}{ccccccc}
 S_\infty'S_\infty & R_\infty'S_\infty\\
 S_\infty'R_\infty & R_\infty'R_\infty
\end{array}
\right), S_2 = Q_{\infty,SR}/(Q_{\infty,S} Q_{\infty,R})^{1/2}.$

By lemma 1, $Q_{\infty,S}$ follows $\chi^2_L$ under $H_0$.  Since $T_{\rm mrAR} \stackrel{p}{\longrightarrow} Q_{\infty,S}$, we have that $P(T_{\rm mrAR} > \chi_L^2(1-\alpha) ) \rightarrow \alpha$. Similarly, since $Q_{\infty,SR}^2/Q_{\infty, R}$ follows $\chi_1^2$ under $H_0$, and  $T_{\rm mrK} \stackrel{p}{\longrightarrow} Q_{\infty,SR}^2/Q_{\infty, R}$, we have that $P(T_{\rm mrK} > \chi_1^2(1-\alpha) ) \rightarrow \alpha$.

For $T_{\rm mrCLR} = \frac{1}{2} (Q_S - Q_R + ((Q_S + Q_R)^2 - 4(Q_SQ_R - Q_{SR}^2))^{1/2})$, let $LR_\infty = \frac{1}{2} (Q_{\infty,S} - Q_{\infty,R} + ((Q_{\infty,S} + Q_{\infty,R})^2 - 4(Q_{\infty,S} Q_{\infty,R} - Q_{\infty,SR}^2))^{1/2}).$ By Lemma 1, $T_{\rm mrCLR} \stackrel{p}{\longrightarrow} LR_\infty$.

Let $w(x; q_r) = P_0(LR_\infty > x|Q_{\infty,R} = q_r) = 1 - \frac{2G\left(\frac{L}{2} \right)}{ \sqrt{\pi} G\left( \frac{L-1}{2} \right) }\int\limits_{0}^{1} {\rm CDF}_{\chi_{L}^2}\left(\frac{x + q_r}{1 + q_r  \frac{z^2}{x}} \right)(1 - z^2)^{\frac{L - 3}{2}} dz,$ where $P_0$ denotes the probability evaluated under $H_0$. We reject $H_0$ if $w(T_{\rm mrCLR}; Q_R)$ is smaller than $\alpha$. Our goal is to show that $\lim\limits_{n_1,n_2 \rightarrow \infty}P(w(T_{\rm mrCLR};Q_R) < \alpha) = \alpha$ when $H_0$ is true.

Denote $F_{0,q_r}(LR_\infty) = w(LR_\infty, q_r)$ as the conditional CDF of $LR_\infty$ given $Q_{\infty, R} = q_r$.
Under $H_0$, we have 
\begin{align*}
\lim\limits_{n_1,n_2 \rightarrow \infty} P(w(T_{\rm mrCLR}; Q_R) < \alpha |Q_{\infty,R} = q_r) &= \lim\limits_{n_1,n_2 \rightarrow \infty}  P_0(w(T_{\rm mrCLR}; q_r) < \alpha |Q_{\infty,R} = q_r)\\
&=  \lim\limits_{n_1,n_2 \rightarrow \infty} P_0(F_{0,q_r}(T_{\rm mrCLR}) < \alpha | Q_{\infty,R} = q_r)\\
& =  P_0(F_{0,q_r}(LR_\infty) < \alpha | Q_{\infty,R} = q_r) = \alpha
\end{align*}
\begin{align*}
\lim\limits_{n_1,n_2 \rightarrow \infty} P_0(w(T_{\rm mrCLR};Q_R) > \alpha) &= \lim\limits_{n_1,n_2 \rightarrow \infty} \int P_0(w(T_{\rm mrCLR};Q_R) > \alpha|Q_{\infty,R} = q_r)f_{Q_{\infty,R}}(q_r) dq_r \\
&= \int \lim\limits_{n_1,n_2 \rightarrow \infty} P_0(p-value > \alpha|Q_{\infty,R} = q_r)f_{Q_{\infty,R}}(q_r) dq_r  \\
&= \int (1 - \alpha)f_{Q_{\infty,R}}(q_r)) dq_r = 1 - \alpha
\end{align*}
where the exchange of the limit and integral in the second last equality is guaranteed by Dominated Convergence Theorem.

\subsection{Proof of Theorem 2.}
Consider estimates obtained from simple linear regression:
\begin{align*}
\widehat{\Gamma_i} &= (Z_{1i}'Z_{1i})^{-1}Z_{1i}'Y_1\\
\widehat{\Sigma}_{\Gamma,ii} &= \frac{Y_1'(I_{n_1} - Z_{1i}(Z_{1i}'Z_{1i})^{-1}Z_{1i}')Y_1}{n_1 Z_{1i}'Z_{1i}}.
\end{align*}
Let  
\[
v_{1i} = \frac{1}{\widehat{\Sigma}_{\Gamma, ii}n_1 + \widehat{\Gamma_i}^2} = \frac{Z_{1i}'Z_{1i}}{Y_1'Y_1},\quad
u_{1i} = v_{1i}\widehat{\Gamma_i} = \frac{Z_{1i}'Y_1}{Y_1'Y_1},
\]
and the ijth entry of $H_1$: $M_{1,ij}\sqrt{v_{1i}v_{1j}} = \frac{Z_{1i}'Z_{1j}}{Y_1'Y_1}$. Thus, $H_1 = \frac{Z_1'Z_1}{Y_1'Y_1}$, and 
\[
\tilde{\Gamma} = H_1^{-1} u_1 = (\frac{Z_1'Z_1}{Y_1'Y_1})^{-1}\frac{Z_1'Y_1}{Y_1'Y_1} 
\]
To adjust for the estimate of standard error, we have 
\[
\tilde{\Sigma}_\Gamma = \frac{1}{n_1-L+1} (1 - \frac{Y_1'Z_1}{Y_1'Y_1}(\frac{Z_1'Z_1}{Y_1'Y_1})^{-1}\frac{Z_1'Y_1}{Y_1'Y_1})(\frac{Z_1'Z_1}{Y_1'Y_1})^{-1}= \frac{1 - u_1'H_1^{-1}u_1}{n_1-L+1}H_1^{-1}
\]
$\tilde{\Gamma}, \tilde{\Sigma}_\Gamma$ are equivalent to estimates obtained from multiple linear regression of $Y_1$ on $Z_1$. Similarly,  
$\tilde{\gamma}, \tilde{\Sigma}_\gamma$ are equivalent to estimates from regressing $D_2$ on $Z_2$. Hence, Theorem 1 can be directly applied here with adjusted estimators $\tilde{\gamma}, \tilde{\Gamma}, \tilde{\Sigma}_\gamma, \tilde{\Sigma}_\Gamma$.

\subsection{Proof of Theorem 3.} 
When $\Sigma_{\Gamma}$ and $\Sigma_{\gamma}$ are diagonal, we have
$$T_{\rm mrAR}(\beta_0) = S^\T (\beta_0)S(\beta_0) =\left(\widehat{\Gamma}  - \beta_0 \widehat{\gamma} \right)^\T \left(\Sphi+ \beta_0^2 \Spi \right)^{-1}\left(\widehat{\Gamma}  - \beta_0 \widehat{\gamma} \right) = \sum\limits_{i = 1}^{L} \frac{(\widehat{\Gamma}_i - \beta_0 \widehat{\gamma}_i)^2}{\widehat{\Sigma}_{ii,\Gamma} + \beta_0^2 \widehat{\Sigma}_{ii,\gamma}} = -2l(\beta_0),$$
where $\widehat{\Sigma}_{ii,\Gamma}$ and $\widehat{\Sigma}_{ii,\gamma}$ are the $i$ diagonal element of $\widehat{\Sigma}_{\gamma}$ and $\widehat{\Sigma}_{\gamma}$, respectively, $l(\beta)$ is the profile log-likelihood function of $\beta$ as in \citet{zhao2020statistical}.

The exact IVW is obtained from minimizing the generalized Q statistic:
\[
\widehat{\beta}_{\rm IVW.exact} = \text{argmin}_{\beta_0} Q_m(w(\beta_0), \beta_0),
\]
where $Q_m(w(\beta_0), \beta_0) = \sum w_i(\beta_0)(\hat{\beta_i} - \beta_0)^2$.

Plugging $w_i(\beta_0) = \frac{\widehat{\gamma_i}^2}{\beta_0^2 Var(\widehat{\gamma_i}) + Var(\widehat{\Gamma_i})}, \hat{\beta}_i = \frac{\widehat{\Gamma_i}}{\widehat{\gamma_i}}$ and $\widehat{\Sigma}_{ii, \gamma}, \widehat{\Sigma}_{ii, \Gamma}$, the estimators of $Var(\widehat{\gamma_i}), Var(\widehat{\Gamma_i})$, into $Q_m(w(\beta_0), \beta_0)$, we have that
\[
Q_m(w(\beta_0), \beta_0) = \sum \frac{\widehat{\gamma_i}^2}{\beta_0^2 \widehat{\Sigma}_{ii, \gamma} + \widehat{\Sigma}_{ii, \Gamma}} (\frac{\widehat{\Gamma_i}}{\widehat{\gamma_i}} - \beta_0)^2 = \sum \frac{(\widehat{\Gamma_i} - \beta_0 \widehat{\gamma_i})^2}{\widehat{\Sigma}_{ii, \Gamma} + \beta_0^2 \widehat{\Sigma}_{ii, \gamma}}
\]

Minimizing $T_{\rm mrAR}$ is equivalent to maximizing $l(\beta_0)$, or minimizing $Q_m(w(\beta_0), \beta_0)$. Thus, $\widehat{\beta}_{\rm mrLIML}$ is equivalent to the profile log likelihood estimator in (3.2) of \citet{zhao2020statistical}, and the exact IVW estimator in Box 2 of \citet{bowden2019improving}.

Furthermore, the confidence interval obtained in Box 3 of \citet{bowden2019improving}:
\[
CI(\hat{\beta}_{IVW.exact}, \alpha) = \{\beta: Q_m(w(\beta_0)), \beta_0 \le \chi^2_{L-1} (1-\alpha)\}
\]
 and the confidence interval obtained from $T_{\rm mrAR}$:
 \[
 CI(T_{\rm mrAR}, \alpha) = \{\beta: T_{\rm mrAR} \le \chi^2_{L} (1-\alpha)\}
 \]
are based on equivalent test statistics ($T_{\rm mrAR} = Q_m(w(\beta_0), \beta_0)$). The only minor difference between those two CIs is that the Q-test uses a critical value of $\chi^2_{L-1}(1 - \alpha)$, while $T_{\rm mrAR}$ uses $\chi^2_{L}(1 - \alpha)$. And this difference is almost negligible when the number of instruments is large.

\subsection{Proof of Approximating Overall F-stat.}
Suppose the individual data $Y_{li}, D_{li}, Z_{li}$ follows the linear structural model (1) 
\begin{align} \label{eq:model_iv}
\begin{split}
&\ Y_{li} = \beta_{\rm int} + D_{li} \beta + \epsilon_{li}, \quad{} D_{li} = \gamma_{\rm int} + Z_{li}^\intercal \gamma+ \delta_{li}, \quad{} E[\epsilon_{li}, \delta_{li} \mid Z_{li}] = 0
\end{split}
\end{align}
Without loss of generality, we assume that $D_{l}, Z_{l}$ are both centralized. For each instrument $Z_{2i}$, the F statistic for each simple linear regression of $D_2$ on $Z_{2i}$ is computed as
\[
F_{SLR,i} = \frac{1}{n_2-L-1}\frac{RSS_0-RSS_i}{RSS_i},
\] where $RSS_i$ denotes residual sum of squares of regressing $D_2$ on $Z_{2i}$, and $RSS_0$ denotes residual sum of squares of regressing $D_2$ on $1$. We have 
\[
RSS_i = D_2'(I_{n_2} - Z_{2i}(Z_{2i}'Z_{2i})^{-1}Z_{2i}')D_2,\quad{}RSS_0 = ||D_2||^2
\]
The F statistic for the multiple linear regression of $D_2$ on $Z_2$ is computed as
\[
F_{MLR} = \frac{L}{n_2-L-1}\frac{RSS_0 - RSS}{RSS}
\]
where $RSS$ denotes residual sum of squares of regressing $D_2$ on $Z_2$, and \begin{align*}
RSS &= D_2'(I_{n_2} - Z_2(Z_2'Z_2)^{-1}Z_2')D_2 \\
&= D_2'(I_{n_2} - Z_2(Z_2'Z_2)Z_2' - \cdots - Z_L(Z_L'Z_L)^{-1}Z_L')D_2\\
&= RSS_1+ \cdots + RSS_L - (L-1)RSS_0
\end{align*}
\begin{align*}
\frac{RSS}{RSS_0} &= \sum \frac{RSS_i}{RSS_0} - (L-1)\\
&= \sum \frac{1}{\frac{F_{SLR,i}}{n_2-L+1}+1}-(L-1)\\
&= 1 - \sum \frac{F_{SLR,i}}{F_{SLR,i} + n_2-L+1}\\
F_{MLR} &= \frac{n_2-L+1}{L}\frac{RSS_0 - RSS}{RSS}\\
&= \frac{n_2-L+1}{L} \frac{\sum \frac{F_{SLR,i}}{F_{SLR,i}+n_2-L+1}}{1 - \sum \frac{F_{SLR,i}}{F_{SLR,i}+n_2-L+1}} \\
&\approx \frac{\sum F_{SLR,i}}{L}
\end{align*}

The approximation follows when $\sum F_{SLR, i} \ll n_2 - L$.

\bibliographystyle{apalike}
\bibliography{paper-ref}